\documentstyle[epsbox,epsfig,12pt]{article}
\newcommand{\fslash}[1]{\mbox{$\!\not\!#1$}}
\newcommand{\bold}[1]{\mbox{\boldmath ${#1}$}}
\newcommand{\agt}{\raisebox{.6ex}{$>$} \hspace{-.8em} \raisebox{-.6ex}
{$\sim$}}

\begin{document}

\baselineskip 4 ex

\title{Quark distributions in nuclear matter and the EMC effect}

\author{H. Mineo \\
Department of Physics, National Taiwan University \\
1 Roosevelt, Section 4, Taipei 10764, Taiwan \\
and \\
Institute of Physics, Academia Sinica \\
Taipei 11529, Taiwan \\
{ } \\
W. Bentz \thanks{Correspondence to: W. Bentz, E-mail: bentz@keyaki.cc.u-tokai.ac.jp}
        \\
       Department of Physics, School of Science, Tokai University \\
       Hiratsuka-shi, Kanagawa 259-1292, Japan \\
{ } \\
       N. Ishii \\
	Department of Physics, Tokyo Institute of Technology,\\
	2-12-1 Ohokayama, meguro, Tokyo 152-8550, Japan\\
{ } \\
       A.W. Thomas \\
       Special Research Centre for the Subatomic Structure of Matter \\
       and \\
       Department of Physics and Mathematical Physics \\
       The University of Adelaide  \\
       Adelaide, SA 5005, Australia \\
{ } \\
       K. Yazaki  \\
       Department of Physics, Tokyo Woman's Christian University \\
       Suginami-ku, Tokyo 167-8585, Japan}

\date{ }
\maketitle
\newpage
\begin{abstract}
Quark light cone momentum distributions in nuclear matter
and the structure function of a bound nucleon are investigated in
the framework of the Nambu-Jona-Lasinio model. This framework describes
the nucleon as a relativistic quark-diquark state, and the nuclear matter
equation of state by using the mean field approximation.
The scalar and vector mean fields in the nuclear medium
couple to the quarks in the nucleon and their effect on the spin
independent nuclear structure function is investigated in detail.
Special emphasis is placed on the important effect of the vector mean field
and on a formulation which guarantees the validity of the number and
momentum sum rules from the outset.

\end{abstract}

\newpage

\section{Introduction}
\setcounter{equation}{0}
The question of how the properties of hadrons bound in the nuclear
medium differ from the properties of
free hadrons is a very important and active field of recent
experimental and theoretical research.
Examples where such medium modifications are important are the
nuclear structure functions
for muon and neutrino scattering \cite{EMCE,EMCT},
proton knock-out processes \cite{PNO},
photoproduction of pions from nuclei \cite{PPP}, etc.
In this work we will be concerned with the
medium modifications of the spin independent nuclear
structure functions measured in deep inelastic
scattering (DIS) of leptons, that is, the EMC
effect~\cite{EMCE}. It is known~\cite{JAF,MIL} that the
Fermi motion and binding effects on the level of nucleons
alone cannot explain the observed reduction of the
nuclear structure function in the range of
Bjorken $x$ between 0.3 and 0.8. As a result, one must face the challenging
task of describing the nuclear systems in terms of
nucleons with internal quark structure which responds to the nuclear
environment and then investigating
whether the binding effects at the quark level can
account for the observations.

At the present time a direct description of many nucleon systems within
QCD itself is out of reach,
and so effective quark theories are the best tools available.
Since we are interested in the effects of the
nuclear medium, these models should account
not only for the properties of the free nucleon, such as
the structure functions, but also for the
saturation properties of nuclear matter (NM). Early work on
this problem~\cite{Thomas:vt}
was based on the quark-meson coupling model.
More recently, using the Nambu-Jona-Lasinio (NJL) model~\cite{NJL} as an
effective chiral quark theory, it has been shown that~\cite{BT,QM}
the quark-diquark description of the single nucleon~\cite{MIN1,MIN2},
which is based on the relativistic Faddeev approach~\cite{FAD},
can be combined successfully with the mean
field description of the NM equation of state (EOS).
The completely covariant description of the
nucleon facilitates the investigation of off-shell effects,
and allows a rigorous incorporation of the Ward identities for baryon 
number and momentum conservation
from the outset. This is particularly important in the description
of the DIS structure functions
\cite{MIN1}\footnote{The model also gives a clear description
of the spontaneous breaking of chiral symmetry in normal matter, and of the
spontaneous breaking of color symmetry and restoration of
chiral symmetry in high density quark matter
\cite{QM}.}.
The model therefore provides a powerful testing ground for investigating
the origin of the EMC effect in terms
of nuclear binding at the quark level. In order to achieve such a
description of nuclear structure
functions, while retaining the simplicity of the NJL model
description, we will restrict our consideration in the present paper
to the case of infinite NM.

The approach which we will follow here is similar
in spirit to the previous investigations
of the EMC effect~\cite{Thomas:vt} using the model of Guichon
and collaborators~\cite{GUI,ST}, which is based on the
MIT bag model description of the single
nucleon. However, one of the
purposes of our present work is to present
a more systematic study of the effects
of the nuclear mean fields, in particular of the vector field,
on the quark light cone (LC) momentum
distributions in NM. It is also important to present a framework which is
consistent with the baryon number and momentum sum rules.
We will, as a first step toward this aim,
restrict ourselves to a valence quark description
of the single nucleon, and to a
mean field description of nuclear matter. Thus, for example,
the effects of the pion
cloud~\cite{Thomas:1983fh,Ericson:1983um,LlewellynSmith:qa,Ericson:1984vt},
which gives rise to the leading non-analytic behaviour of the parton
distribution functions as a function of quark mass~\cite{Leinweber:2001ui},
will not be
discussed here. It has, however, been included in the NJL model
description of the free nucleon structure functions \cite{MIN1} and will be
incorporated in the nuclear calculations in future work.

The outline of this paper is as follows. In Sect. 2 we
discuss the LC momentum distributions
of quarks in NM by using the convolution formalism.
Our main task is to explain
the important effect of the mean vector field in a
manner consistent with the sum rules.
In Sect. 3 we introduce the model used to describe both the single
nucleon and the EOS of NM, while
in Sect. 4 we present the numerical results and discussions
concerning the origin of the EMC effect
in our description. A summary and conclusions are given in Sect. 5.

\section{Light cone momentum distributions}
\setcounter{equation}{0}
The LC momentum distribution of quarks per nucleon in a nucleus with mass number $A$ is
defined as
\begin{equation}
f_{q/A}(x_A)=\frac{P_-}{A^2} \int \frac{{\rm d}w^-}{2\pi} e^{i P_- x_A w^-/A}
<A,P|\overline{\psi}(0) \gamma^+ \psi(w^-)|A,P>\,.  \label{fqa}
\end{equation}
Here $P^{\mu}$ is the total 4-momentum of the nucleus
and $P_-$ its ``minus component
\footnote{Our notations for LC variables are $a^{\pm}=(a^0\pm a^3)/\sqrt{2}$,
$a_{\pm}=(a_0\pm a_3)/\sqrt{2}$. We will frequently call $p_-$ ($p_+$) the ``minus
(plus) component'' of the 4-vector $p^{\mu}$.}, and $\psi$ is the quark field.
Because we consider only the isoscalar case, $N=Z$, without the Coulomb
force, we need only the isospin
symmetric combination, $f_{q/A} \equiv f_{u/A}+f_{d/A}$.

Equation~(\ref{fqa}) corresponds to the quark 2-point function in the nucleus,
where the minus component of the quark momentum
is fixed as $k_-=P_- x_A/A$, and the other LC
components are integrated out.
The variable $x_A$ is therefore $A$ times the fraction
of the minus component of the total
momentum carried by a quark: $x_A = A\,k_-/P_-.$
We will work in the rest frame of the nucleus,
where $P^{\mu}=(M_A,{\bold 0})$, and
\begin{eqnarray}
P_-/A = \overline{M}_N/\sqrt{2}\,
\label{mbar}
\end{eqnarray}
where $\overline{M}_N=M_A/A$ is the mass per nucleon. In this frame we have
$x_A= \sqrt{2} k_-/\overline{M}_N$, or
\begin{eqnarray}
x_A = \frac{M_{N0}}{\overline{M}_N}\,x\,,
\label{bja}
\end{eqnarray}
where $x$ is the ordinary Bjorken variable for DIS on a free nucleon
with mass~\footnote{We will use the subscript $0$ to
denote masses at zero baryon density.} $M_{N0}=940$ MeV.

Since in the nuclear medium there is no
Lorentz invariance with respect to the motion of a single nucleon,
it is natural to use
a non-covariant normalization of state vectors. In this work we use
the ``non-covariant LC normalization'', where
\begin{equation}
<A,P|\overline{\psi} \gamma^+ \psi|A,P> =  3\,A .   \label{norma}
\end{equation}
In this normalization, the vertex function of a
bound state, like the nucleon, is defined
by the residue of the $t$-matrix in the corresponding
channel at the pole in the plus
component of the momentum of the bound state --
see Subsect.~3.1 for the case of the nucleon.

In this Section we wish to discuss some features of the
distribution (\ref{fqa})
which are largely independent of the details of the model for
the nucleon and NM.
There are two approximations which we need for this purpose.
The first, which corresponds to the mean field approximation, is that
in the presence of the nuclear scalar and vector mean fields the
energy spectrum of a single nucleon with 3-momentum ${\bold p}$ has the form
\begin{eqnarray}
p_0 = \epsilon_p = \sqrt{M_N^2+{\bold p}^2} + 3\,V_0 \equiv E_p +
3 \, V_0\,.
\label{spec}
\end{eqnarray}
Here the effective nucleon mass, $M_N=M_N(M)$, is obtained as
a function of the effective
(constituent) quark
mass, $M$, by solving the bound state equation for the nucleon using
the solution of the in-medium gap equation for $M$.
We denote by $V^{\mu}=(V_0, {\bold 0})$ the
constant mean vector field acting on a quark in the nuclear
medium. That is, the quark
Hamiltonian in the mean field approximation for NM at rest has the form
\begin{eqnarray}
\hat{H}_q = \hat{h}_q + V_0\,\hat{Q}, \label{hq}
\end{eqnarray}
where $\hat{h}_q$ is the quark Hamiltonian without the mean vector field, and
\\ ${\displaystyle \hat{Q} = \int {\rm d}^3x\,\psi^{\dagger}(x) \psi(x)}$
is the
quark number operator.
We also define the ``kinetic momentum'' of the nucleon in the
medium as
\begin{eqnarray}
p_{\rm N}^{\mu}= p^{\mu} - 3\,V^{\mu}\,.  \label{kin}
\end{eqnarray}

The second assumption is that the distribution (\ref{fqa})
can be evaluated by using the
familiar one-dimensional convolution formula
\begin{equation}
f_{q/A}(x_A) = \int {\rm d}y_A \int {\rm d}z \,\delta(x_A-y_A z)\,f_{q/N}(z)\,f_{N/A}(y_A) \, ,
\label{conv}
\end{equation}
where $f_{q/N}(z)$ and $f_{N/A}(y_A)$ are the LC
momentum distributions of quarks
in the nucleon, and of nucleons in the nucleus (per nucleon), respectively:
\begin{eqnarray}
f_{q/N}(z) &=& p_-  \int \frac{{\rm d}w^-}{2\pi} e^{i p_- z w^-}
<N,p|\overline{\psi}(0) \gamma^+ \psi(w^-)|N,p>,  \label{fqn} \\
f_{N/A}(y_A) &=& \frac{P_-}{A^2}
\int \frac{{\rm d}w^-}{2\pi} e^{i P_- y_A w^-/A}
<A,P|\overline{\psi}_N(0) \gamma^+ \psi_N(w^-)|A,P>\,.
\nonumber \\  \label{fna}
\end{eqnarray}
Here $\psi_N$ is the nucleon field~\footnote{In our quark-diquark
model for the nucleon, which
will be discussed in Subsect.3.1, the interpolating field for the nucleon
is local --
see Ref.~\cite{QM} for the explicit expression.
We also note that the main assumption which leads
to the convolution formula (\ref{conv}) is that
the LC momentum
distribution of a nucleon with virtuality $\ell^2$ has a sharp
peak at $\ell^2=M_N^2$, so that the quark momentum distribution in the nucleon
can be evaluated at $\ell^2=M_N^2$ and taken outside
the integral over the virtuality.},
and the normalizations are
\begin{eqnarray}
<N,p|\overline{\psi} \gamma^+ \psi|N,p> &=& 3 \,,   \label{norm1} \\
<A,P|\overline{\psi}_N \gamma^+ \psi_N|A,P> &=& \frac{1}{\sqrt{2}}
<A,P|\psi_N^{\dagger} \psi_N |A,P> = A \,,   \label{norm2}
\end{eqnarray}
where the first equality in Eq.~(\ref{norm2}) holds in the rest system of NM.
It follows from the expressions (\ref{fqn}) and (\ref{fna})
that $z$ is the fraction of the minus
component of the nucleon momentum ($p^{\mu}$)
carried by a quark, $z=k_-/p_-$, and
$y_A$ is $A$ times the fraction
of the minus component of the nuclear momentum ($P^{\mu}$)
carried by a nucleon,
$y_A=A\,p_-/P_- = \sqrt{2} p_-/\overline{M}_N$.

In the following two Subsections we will
discuss the distributions $f_{q/N}(z)$ and $f_{N/A}(y_A)$
separately. In particular, we focus on their dependence
on the mean vector field and the validity of the baryon
number and momentum sum rules. In Subsect. 2.3 we will
return to the convolution formula (\ref{conv}).

\subsection{Quark LC momentum distribution in the nucleon}
If one has a model to describe the nucleon as a bound state of quarks,
the evaluation
of the distribution $f_{q/N}(z)$ of Eq.(\ref{fqn}) can be reduced
to a straightforward Feynman diagram
calculation by noting that it can be expressed as follows \cite{JAF,MIN1}
\footnote{We remind the reader that for connected LC correlation
functions the $T$-product is identical to the usual product \cite{JAFNP}.}:
\begin{eqnarray}
f_{q/N}(z) = -i \int \frac{{\rm d}^4 k}{(2\pi)^4}\,\delta\left(z-\frac{k_-}{p_-}\right)\,
{\rm tr}\left(\gamma^+ M(p,k)\right)\,,
\label{fey1}
\end{eqnarray}
where the quark two-point function in the nucleon is given by
\begin{eqnarray}
M_{\beta \alpha}(p,k)=i \int {\rm d}^4w\,e^{i k\cdot w}\, \langle N,p| T\,\left[\overline{\psi}_{\alpha}(0)
\,\psi_{\beta}(w)\right]|N,p\rangle\,,
\label{p1}
\end{eqnarray}
and the trace in (\ref{fey1}) refers to the Dirac and isospin indices.
In the quark-diquark model, which will be introduced in Sect.3,
such a direct calculation is based on
the Feynman diagrams of Fig.1 -- as discussed in Appendix A.

\begin{figure}[h]
\begin{center}
\epsfig{file=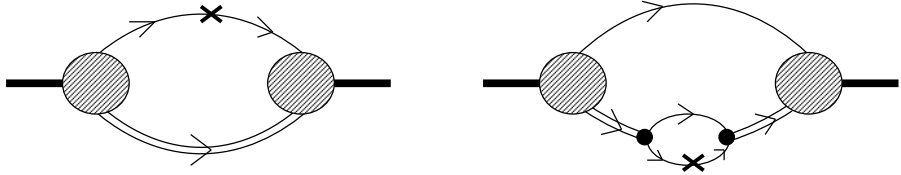,angle=0,width=12cm}
\caption{Graphical representation of
the function $f_{q/N}(z)$ of Eq.(\ref{fey1}) in the
quark-diquark model.
The single (double) line denotes the quark propagator (scalar
diquark t-matrix), the hatched circle
stands for the quark-diquark vertex function and
the operator insertion, denoted by a cross,
stands for $\gamma^+ \delta\left(z-k_-/p_-\right)$.}
\end{center}
\end{figure}

The difference with respect to the free nucleon case considered in
Ref.~\cite{MIN1} is that all propagators have to
be replaced by those including the scalar and vector
mean fields. The effect of the scalar
field can easily be incorporated into the
effective masses, but the effect of the
vector field is a bit more subtle. In this Subsection we wish to
determine this dependence on the vector field more generally,
without explicit reference to
{}Feynman diagrams.

{}For this purpose, we note the following two points.
{}First, the quark field satisfies
\begin{equation}
\psi(w) = e^{i {\hat P}_q\cdot w} \psi(0) e^{-i {\hat P}_q\cdot w}\,,  \label{heis}
\end{equation}
where ${\hat P}_q^{\mu}=({\hat H}_q,
{\hat {\bold P}}_q)$ the 4-momentum operator for quarks.
The dependence of the quark Hamiltonian on the
vector field has been given  in Eq.~(\ref{hq}). Since
$[\hat{h}_q,\hat{Q}]=0$ and $\hat{Q}$ is time independent, we have
\begin{eqnarray}
\frac{\partial \psi(w)}{\partial V_0} =
i w^0 [{\hat Q}, \psi(w)] = -i w^0 \psi(w),
\nonumber
\end{eqnarray}
the solution of which can be written in the form of a
local gauge transformation
\footnote{To check the relation (\ref{gauge}) in classical field
theory, it is sufficient to
note that $\overline{\psi}_{0} \left(i\fslash{\partial} - M \right)\psi_{0} =
\overline{\psi} \left(i\fslash{\partial} - \fslash{V} - M \right)\psi$,
i.e., the local
gauge transformation generates the coupling of the quarks
to the (constant) vector potential.}:
\begin{equation}
\psi(w) = e^{-i V \cdot w} \psi_{0}(w).  \label{gauge}
\end{equation}
Here $\psi_{0}$ is the quark field in the absence of the vector potential.

Second, if the nucleon state $|N,p \rangle$ is an
eigenstate of the quark Hamiltonian, ${\hat H}_q$,
with eigenvalue $p^0$ given by Eq.(\ref{spec}),
and of the quark number operator ${\hat Q}$
with eigenvalue $3$, it is also an eigenstate of ${\hat h}_q$
with eigenvalue $p_{\rm N}^0$,
which is related to $p^0$ by Eq.(\ref{kin}).
We can express this fact in the following way:
\begin{equation}
|N, p \rangle = |N,p_{\rm N} \rangle_0 \,,  \label{state}
\end{equation}
where the subscript $0$ characterizes a nucleon
state without the external vector field, which we normalize in
the same way as Eq.~(\ref{norm1}).
If we insert
the relations (\ref{gauge}) and (\ref{state}) into Eq.(\ref{fqn}), we obtain
\begin{eqnarray}
f_{q/N}(z)= \frac{p_-}{p_{{\rm N}-}}\,f_{q/N0}(z'\equiv \frac{p_-}{p_{{\rm N}-}}z - \frac{V_-}{p_{{\rm N}-}})\,,
\label{relation}
\end{eqnarray}
where $f_{q/N0}(z')$ is the distribution function in
the absence of the vector potential
and can simply be evaluated by following
the method for the free nucleon case \cite{MIN1} with
modified masses.
Since in our valence quark picture (no sea quarks) the function
$f_{q/N0}(z')$ has support for $0<z'<1$ \cite{MIN1,JAFNP},
the distribution $f_{q/N}(z)$ has support in the
region $R_N(z)$ defined by
\begin{equation}
R_N(z): \,\,\,\,\,\,\,\,\,\,\,\,\,\,\,\,\,\,\,\,
\frac{V_-}{p_-} < z < 1 - \frac{2 V_-}{p_-}  \,\,\,\,\,\,\,\,\,\,\,\,\,\,\,
({\rm support}\,\,{\rm of}\,\,f_{q/N}(z))\,. \label{supq}
\end{equation}

Here we make several comments. First, in the quark-diquark
model, the result (\ref{relation})
also follows from a straightforward calculation of the Feynman
diagrams of Fig.1 -- as shown in Appendix A.
Second, this discussion concerning the dependence on the
vector field can easily be generalized
to any hadron (H) with quark number $H$. This leads to
\begin{eqnarray}
f_{q/H}(z)= \frac{p_-}{p_{{\rm H}-}}\,f_{q/H0}(z'\equiv
\frac{p_-}{p_{{\rm H}-}}z - \frac{V_-}{p_{{\rm H}-}})\,,
\label{relation1}
\end{eqnarray}
where $p^{\mu}$ is the momentum of the hadron and
\begin{eqnarray}
p_{\rm H}^{\mu}=p^{\mu}-H\,V^{\mu} \label{kinh}
\end{eqnarray}
is its kinetic momentum.
The distribution (\ref{relation1}) has support in the
region $R_H(z)$ defined by
\begin{equation}
R_H(z): \,\,\,\,\,\,\,\,\,\,\,\,\,\,\,\,\,\,\,\,
\frac{V_-}{p_-} < z < 1 - \frac{(H-1) V_-}{p_-}  \,\,\,\,\,\,\,\,\,\,\,\,\,\,\,
({\rm support}\,\,{\rm of}\,\,f_{q/H}(z))\,.
\label{supq1}
\end{equation}
The quark number, $H=3$, corresponds to the case of the
nucleon (H=N), which was discussed above. Third,
the relation (\ref{relation}) shows that $f_{q/N}(z)$ actually
depends not only on $z$, but also on
$p_-$, or on $y_A$
(since $y_A=Ap_-/P_-$), although we do not indicate this fact
explicitly in our notation.
This dependence on the nucleon momentum, which arises
because in the medium there is no Lorentz
invariance with respect to the motion of a single nucleon,
has to be taken into account
in the convolution integral (\ref{conv}), as will be discussed in Sect.3.
Fourth, in Appendix B we show the consistency of (\ref{supq1}) with the
spectral representation
\begin{eqnarray}
f_{q/H}(z) = p_-\,\sqrt{2} \sum_n \delta(p_-(1-z)-p^{(n)}_-)\,\,
|\langle n|\psi_{(+)}|H,p \rangle |^2\,,  \label{spq}
\end{eqnarray}
where $\psi_{(+)}= \Lambda_+\,\psi$ with
$\Lambda_+=\gamma^0 \gamma^+ /\sqrt{2}$.
That is, following the discussions in
Ref.\cite{JAFNP}, there are contributions from connected and
semi-connected diagrams to the r.h.s. of Eq.(\ref{spq}) and the
connected piece (diagram 1 of Fig. 8 in Appendix B) has
support only in the region (\ref{supq1}),
while the semi-connected pieces have support in different regions of $z$.
The connected piece can be interpreted as the LC momentum distribution
of quarks in the hadron H
in the sense of the parton model.

If we suppose that the distribution $f_{q/H0}(z)$ satisfies the
number and momentum
sum rules, which is the case in the quark-diquark model for the
nucleon ($H=N$) \cite{MIN1},
Eq.(\ref{relation1}) can be used to
confirm the validity of these sum rules for $f_{q/H}(z)$:
\begin{eqnarray}
\int_{R_H} {\rm d}z\,f_{q/H}(z) &=&
\int_0^1 {\rm d}z'\, f_{q/H0}(z') = H\,, \label{ns} \\
\int_{R_H} {\rm d}z\,z\,f_{q/H}(z) &=&
\frac{p_{{\rm H}-}}{p_-} \int_0^1 {\rm d}z'\, z'\, f_{q/H0}(z')
+ \frac{V_-}{p_-} \int_0^1 {\rm d}z'\, f_{q/H0}(z')  \nonumber \\
&=& \frac{p_{{\rm H}-}}{p_-} + \frac{H V_-}{p_-} = 1\,. \label{ms}
\end{eqnarray}
This derivation clearly shows that the shift of
the momentum fraction induced by the vector potential
(see Eq.(\ref{relation1})) is
necessary to satisfy the number and momentum sum rules at the same time.

\subsection{Nucleon LC momentum distribution in nuclear matter}
The function $f_{N/A}(y_A)$ of Eq.(\ref{fna})
has been discussed in detail in Ref.\cite{MIL}
for the case of a mean field description of NM and we briefly
summarize the derivation here for convenience.
First we express (\ref{fna}) as follows:
\begin{eqnarray}
f_{N/A}(y_A) = \frac{-i}{A} \int \frac{{\rm d}^4 p}{(2\pi)^4}\,
\delta\left(y_A-\frac{\sqrt{2} p_-}{\overline{M}_N}\right)\,
{\rm tr}\left(\gamma^+ G_N(p)\right)\,, \label{fey2}
\end{eqnarray}
where we used Eq.(\ref{mbar}) and introduced the nucleon
two-point function in the medium
\begin{eqnarray}
G_{N,\beta \alpha}(p)=i \int {\rm d}^4 w\,e^{i p\cdot w}\, \langle A,P| T\,\left[\overline{\psi}_{N,\alpha}(0)\,
\psi_{N,\beta}(w)\right]|A,P\rangle\,.
\label{p2}
\end{eqnarray}
The relation to the familiar Feynman propagator in
the medium $S_N(p)$ \cite{SW} is given by
\footnote{To verify Eq.(\ref{gn}), we note that the
usual (dimensionless) NM ground state
$|\rho>$ at the saturation density $\rho$ satisfies
$\langle \rho|\psi_N^{\dagger} \psi_N|\rho\rangle=\rho$.
Comparison with the normalization (\ref{norm2})
gives $|A,P\rangle = \sqrt{V}\,2^{1/4} |\rho\rangle$,
which leads to Eq.(\ref{gn}).}
\begin{eqnarray}
G_N(p) = V \sqrt{2}\, S_N(p)\,, \label{gn}
\end{eqnarray}
where $V$ is the volume of the system.
The propagator $S_N(p)$ consists of a ``Feynman part''
$S_{NF}(p)=\left(\fslash{p}_{\rm N}-M_N +i\epsilon \right)^{-1}$, and
a density dependent part, $S_{ND}(p)$,
which takes into account the fact that NM consists of nucleons
with momenta up to the
Fermi momentum $p_F$:
\begin{eqnarray}
S_{ND}(p) = \frac{\fslash{p}_{\rm N} + M_N}{E_{p}}\,i\pi\,\delta\left(p_{{\rm N}0}-E_{p}\right)\,
\Theta(p_F-|{\bold p}|)\, ,
\label{snd}
\end{eqnarray}
where the kinetic momentum $p_{\rm N}$ in these expressions
was defined in Eq.(\ref{kin}).
In the mean field description of the NM ground state,
we have to replace \cite{SW}
$S_N \rightarrow S_{ND}$ in loop integrals like
in Eq.(\ref{fey2}), which becomes
\begin{eqnarray}
f_{N/A}(y_A) = \frac{-i \sqrt{2}}{\rho} \int \frac{{\rm d}^4 p}{(2\pi)^4}\,
\delta\left(y_A-\frac{\sqrt{2} p_-}{\epsilon_F}\right)\,
{\rm tr}\left(\gamma^+ S_{ND}(p)\right)\,. \label{fey3}
\end{eqnarray}
Here $\rho$ is the baryon density and we used the fact
that at the saturation density the mass per nucleon is equivalent to the
{}Fermi energy~\footnote{This is the ``Hugenholtz-van Hove theorem'',
which follows simply from the definition of the chemical potential
($\epsilon_F=\partial {\cal E}/\partial \rho$,
where ${\cal E}$ is the energy density of NM) and the saturation
condition for $\overline{M}_N = {\cal E}/\rho$:
\begin{eqnarray}
\frac{\partial}{\partial \rho} \frac{\cal E}{\rho} =
\frac{{\epsilon}_F}{\rho} - \frac{\cal E}{\rho^2} = 0
\Rightarrow \epsilon_F = \overline{M}_N. \nonumber
\end{eqnarray}}
that is, using Eq.(\ref{spec}),
\begin{eqnarray}
\overline{M}_N = \epsilon_F = \sqrt{p_F^2+M_N^2}+
3V_0 \equiv E_F + 3 V_0. \label{hh}
\end{eqnarray}
The expression (\ref{fey3}), which is graphically
represented by Fig.2, can easily be
evaluated with the result
\begin{eqnarray}
\lefteqn{f_{N/A}(y_A) = \frac{3}{4} \left(\frac{\epsilon_F}{p_F}\right)^3
\left[ \left(\frac{p_F}{\epsilon_F}\right)^2
-(1-y_A)^2\right]}    \label{fn1} \\
& & 1-\frac{p_F}{\epsilon_F} < y_A < 1 + \frac{p_F}{\epsilon_F}. \label{rya}
\end{eqnarray}
\begin{figure}[h]
\begin{center}
\epsfig{file=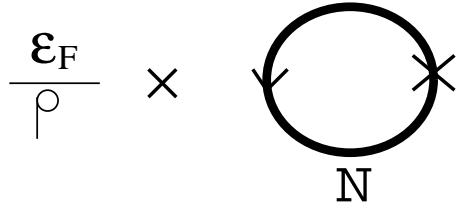,angle=0,width=6cm}
\caption{Graphical representation of
the function $f_{N/A}(y_A)$ of Eq.(\ref{fey3}).
The closed line denotes the nucleon propagator $S_{ND}$, and
the operator insertion, denoted by a cross, stands for
$\gamma^+ \delta\left(p_- - \frac{\epsilon_F\,y_A}{\sqrt{2}}\right)$.}
\end{center}
\end{figure}
It is easy to verify that (\ref{fn1}) satisfies the
relation~\footnote{The relation (\ref{relation0}) is formally
similar to (\ref{relation1}), which refers to
a hadron in an external vector field.
However, one should not push this analogy too far,
since no stable NM state exists without its internal vector field and therefore
it does not make sense to define the quantity
$f_{N/A0}(y'_A)$ simply by setting $V_0=0$
directly in Eq.(\ref{fna}).}
\begin{equation}
f_{N/A}(y_A) = \frac{\epsilon_F}{E_F} f_{N/A0}\left(y'_A \equiv
\frac{\epsilon_F}{E_F}y_A - \frac{3 V_0}{E_F}\right)\,,
\label{relation0}
\end{equation}
where $f_{N/A0}(y'_A)$ is the distribution obtained by
setting $V_0=0$ in Eqs.~(\ref{fn1}) and (\ref{rya}):
\begin{eqnarray}
\lefteqn{f_{N/A0}(y'_A) = \frac{3}{4} \left(\frac{E_F}{p_F}\right)^3
\left[ \left(\frac{p_F}{E_F}\right)^2 -(1-y'_A)^2\right]}   \label{fn0} \\
& & 1-\frac{p_F}{E_F} < y'_A < 1 + \frac{p_F}{E_F} \label{rya0}.
\end{eqnarray}

The number and momentum sum rules for the nucleons are
easily verified by noting that
any function of the form ${\displaystyle f(y) = \frac{3}{4} \frac{1}{a^3}
\left[a^2 - (1-y)^2 \right]}$, where $1-a < y < 1+a$, satisfies
${\displaystyle \int {\rm d} y \, f(y) = \int {\rm d} y\,y\, f(y) = 1}$
for {\em any} value of $a$.
That is, both $f_{N/A}$ and $f_{N/A0}$ satisfy
the baryon number and momentum sum rules
\begin{eqnarray}
\int {\rm d} y_A \, f_{N/A}(y_A) &=& \int {\rm d} y_A\,y_A\, f_{N/A}(y_A) = 1\,, \label{sumn0} \\
\int {\rm d} y'_A \, f_{N/A0}(y'_A) &=& \int {\rm d} y'_A\,y'_A\, f_{N/A0}(y'_A) = 1\,, \label{sumn}
\end{eqnarray}
where the ranges of integration are given by Eqs.~(\ref{rya}) and (\ref{rya0}).

Although the form (\ref{fn1}) formally satisfies
the sum rules for any values of
$p_F$ and $\epsilon_F$, one should keep in mind
that it is valid only at the saturation point --
since we used the relation (\ref{hh}) which follows from
the saturation condition. Neither the
momentum sum rule, nor the relation (\ref{relation0}),
would hold if one were to take
$\overline{M}_N$ to be a fixed parameter.

\subsection{Quark momentum distribution in nuclear matter}
In order to obtain the dependence of the quark LC momentum
distributions in NM on the vector
field, we have to insert (\ref{relation}) and
(\ref{relation0}) into the convolution formula
(\ref{conv}).
As we pointed out already, because of the presence of the
mean vector field, which refers to the NM
rest frame, $f_{q/N}(z)$ actually depends also on $y_A$.
Convolution integrals of the type (\ref{conv})
are treated generally in Appendix C and the result is as follows:
\begin{eqnarray}
f_{q/A}(x_A) = \frac{\epsilon_F}{E_F} f_{q/A0}(x'_A = \frac{\epsilon_F}{E_F} x_A - \frac{V_0}{E_F}).
\label{fqa1}
\end{eqnarray}
Here the distribution without the explicit effect of the
mean vector field is denoted by $f_{q/A0}(x'_A)$ and defined as
\begin{equation}
f_{q/A0}(x'_A) =
\int {\rm d}y'_A \int {\rm d}z' \,\delta(x'_A-y'_A z')\,f_{q/N0}(z') \,
f_{N/A0}(y'_A) \,.
\label{conv0}
\end{equation}
The distribution $f_{q/A0}(x'_A)$ includes,
besides the Fermi motion of nucleons,
only the effect of the scalar field,
which is incorporated easily into the effective
masses. Because of the constraint $0<z'<1$ and the support
of $y_A'$ given in Eq.~(\ref{rya0}), the distribution
(\ref{conv0}) has support for $0<x_A' < 1+p_F/E_F$
and the distribution (\ref{fqa1}), which
includes the effect of the mean vector field, has support in the region
\begin{equation}
R_A(x_A): \,\,\,\,\,\,\,\,\,\,\,\,\,\,\,\,\,\,\,\,
\frac{V_0}{\epsilon_F} < x_A < \frac{E_F+p_F+V_0}{\epsilon_F}  \,\,\,\,\,\,\,\,\,\,\,\,\,\,\,
({\rm support}\,\,{\rm of}\,\,f_{q/A}(x_A))\,. \label{supqa}
\end{equation}

Since we have already shown the sum rules for $f_{q/N}$ and
$f_{N/A}$, the sum rules for $f_{q/A}(x_A)$ follow immediately from the
convolution formula (\ref{conv}).
One can also make direct use of Eq.~(\ref{fqa1}) to
check the validity of the sum rules,
which shows the importance of the shift arising from the
mean vector field more explicitly. Indeed, using the fact that
the distribution (\ref{conv0}) satisfies the
number and momentum sum rules, we obtain
\begin{eqnarray}
\int_{R_A} {\rm d} x_A \,f_{q/A}(x_A) &=&
\int_0^{1+a_0} {\rm d} x_A' \,f_{q/A0}(x_A') = 3 \label{nsr}\,, \\
\int_{R_A} {\rm d} x_A \, x_A \, f_{q/A}(x_A) &=&
\frac{E_F}{\epsilon_F} \int_0^{1+a_0} {\rm d} x_A'\,x_A'\, f_{q/A0}(x_A')
+ \frac{V_0}{\epsilon_F} \int_0^{1+a_0} {\rm d} x_A' \, f_{q/A0}(x_A') \nonumber \\
&=& \frac{E_F}{\epsilon_F} + \frac{3 V_0}{\epsilon_F} = 1\,,   \label{msr}
\end{eqnarray}
where ${\displaystyle a_0=\frac{p_F}{E_F}}$.

\section{Model for the nucleon and nuclear matter}
\setcounter{equation}{0}
In order to evaluate the quark distributions and
structure functions in NM, we need a model
for the nucleon and the EOS of NM. In this work, we use the NJL
model as an effective quark
theory to describe the nucleon as a quark-diquark bound state and NM in the
mean field approximation. The details are explained in Refs.~\cite{BT,QM},
and here we will only briefly summarize those points which will be
needed for our calculations.

The NJL model is characterized by a chirally
symmetric 4-fermi interaction between the
quarks. Any such interaction can be Fierz
symmetrized and decomposed into various
$q\overline{q}$ channels \cite{FAD}. Writing out explicitly only those channels which are relevant
for our present discussion, we have
\begin{equation}
{\cal L}={\overline \psi}\left(i \fslash{\partial}-m\right) \psi + G_{\pi}
\left( \left(\overline{\psi}\psi \right)^2
	- \left(\overline{\psi}(\gamma_5\bold{\tau})\psi\right)^2 \right)
- G_{\omega} \left( \overline{\psi} \gamma^{\mu} \psi \right)^2  + \ldots
\label{lag}
\end{equation}
where $m$ is the current quark mass.
The quark bilinears
$\overline{\psi}\psi$ and $\overline{\psi}\gamma^{\mu}\psi$ have expectation
values in the NM ground state $|\rho \rangle$, which we will separate as usual according to
$\overline{\psi}\psi=
\langle \rho|\overline{\psi}\psi|\rho \rangle + (:\overline{\psi}\psi:)$
and $\overline{\psi}\gamma^{\mu}\psi=
\langle \rho|\overline{\psi}\gamma^{\mu}\psi|\rho \rangle
+ (:\overline{\psi}\gamma^{\mu}\psi:)$.
The Lagrangian can then be expressed as
\begin{eqnarray}
{\cal L}={\overline \psi}\left(i \fslash{\partial}- M - \fslash{V}
\right) \psi - \frac{(M-m)^2}{4 G_{\pi}} + \frac{V_{\mu} V^{\mu}}{4G_{\omega}} + {\cal L}_I,
\label{lag1}
\end{eqnarray}
where $M=m-2G_{\pi} \langle \rho|\overline{\psi}\psi|\rho \rangle$ and
$V^{\mu}= 2 G_{\omega} \langle \rho|\overline{\psi}\gamma^{\mu}\psi|\rho \rangle$, and ${\cal L}_I$
is the normal ordered interaction Lagrangian,
including the counter terms linear in the
normal ordered products.

In order to fully define the model one has to define a cut-off procedure.
In all calculations in this paper
we will use the proper time regularization
scheme~\cite{IR,BT}, where one evaluates
loop integrals over a product of propagators by
introducing Feynman parameters, performing the Wick rotation and
replacing the denominator ($A$) of the loop integral according to
\begin{eqnarray}
\frac{1}{A^n} \rightarrow \frac{1}{(n-1)!} \int_{1/\Lambda_{\rm UV}^2}^{1/\Lambda_{\rm IR}^2}
{\rm d}\tau \, \tau^{n-1}\,e^{-\tau A}\,\,\,\,\,\,\,\,\,(n\geq 1), \label{pt}
\end{eqnarray}
where $\Lambda_{\rm IR}$ and $\Lambda_{\rm UV}$ are the infrared (IR)
and  ultraviolet (UV) cut-offs, respectively.
Although there are no IR divergences in the theory,
the IR cut-off plays the important role of
eliminating the imaginary part of the loop integrals.
That is, it eliminates the
unphysical thresholds for the decay of the nucleon and
mesons into quarks \cite{IR}, thereby taking into consideration a particular
aspect of confinement physics.
It has been shown in Ref.\cite{BT} that this is crucial
to describe saturation of the NM binding energy in the NJL model.

\subsection{Model for the nucleon}

One can use a further Fierz transformation to
decompose ${\cal L}_I$ into a sum of
$qq$ channel interaction terms \cite{FAD}. For
our purposes we need only the interaction in the scalar diquark
($J^{\pi}=0^+, T=0$, color ${\overline 3}$) channel:
\begin{eqnarray}
{\cal L}_{I,s} = G_s \left(\overline{\psi}\left(\gamma_5 C\right)\tau_2
\beta^A \overline{\psi}^T\right) \left(\psi^T\left(C^{-1}\gamma_5\right)
\tau_2 \beta^A \psi\right),
\label{lags}
\end{eqnarray}
where $\beta^A=\sqrt{3/2}\,\, \lambda^A \,\,(A=2,5,7)$ are the color
${\overline 3}$ matrices and $C=i\gamma_2 \gamma_0$.
The coupling constant $G_s$ will be
determined so as to reproduce the free nucleon mass.

The reduced t-matrix in the scalar diquark channel is given by \cite{BT}
\begin{eqnarray}
\tau(q)=\frac{4iG_s}{1 + 2 G_s \Pi_{s}(q)} = \tau_0(q_{\rm D})
\label{taus}
\end{eqnarray}
with the scalar $qq$ bubble graph
\begin{eqnarray}
\Pi_{s}(q)=6 i \int \frac{{\rm d}^4 k}{(2\pi)^4} tr_D \left[
	\gamma_5 S(k) \gamma_5  S \left(-(q-k)\right) \right]
= \Pi_{s0}(q_{\rm D})\,.
\label{bubbs}
\end{eqnarray}

In the second equalities of Eqs.~(\ref{taus}) and (\ref{bubbs})
we indicated that these
quantities can be obtained from their expressions without the
effect of the mean vector field by replacing
$q\rightarrow q_{\rm D}$,
where $q_{\rm D}$ is the kinetic momentum of the diquark
defined by $H=2$ in (\ref{kinh}). This is
easily seen by using the following form of the quark propagator
in the presence of the mean vector field:
\begin{eqnarray}
S(k) =\left(\fslash{k}_{\rm Q} - M +i\epsilon \right)^{-1}=S_0(k_{\rm Q})\,,
\label{qf}
\end{eqnarray}
where the kinetic momentum of the quark is defined by $H=1$ in (\ref{kinh}) and
$S_0$ is the usual Feynman propagator without the effect of the
external vector field.

The relativistic Faddeev equation in the NJL model has been solved numerically
for the free nucleon in earlier work~\cite{FAD}. For
the present purposes, however, because we need to solve the equation many
times in order to obtain self-consistency in-medium,
we restrict ourselves to the static approximation to the Faddeev equation
\cite{STAT}, where the momentum dependence of the quark exchange kernel
is neglected. The solution for the quark-diquark t-matrix
then takes the simple analytic form
\begin{eqnarray}
T(p)=\frac{3}{M} \frac{1}{1+\frac{3}{M} \Pi_{N}(p)} = T_0(p_{\rm N})\,,
\label{tn}
\end{eqnarray}
with the quark-diquark bubble graph given by
\begin{eqnarray}
\Pi_{N}(p)=-\int \frac{{\rm d}^4 k}{(2\pi)^4}\, S(k)\, \tau(p-k) = \Pi_{N0}(p_{\rm N}).
\label{bubbn}
\end{eqnarray}
In these relations, again, the subscript
$0$ characterizes the quantities without the effect of the mean vector field.
The nucleon mass $M_{N}$ is defined as the
pole of (\ref{tn}) at $\fslash{p}_{\rm N}=M_N$
and the positive energy spectrum has the form given in Eq.(\ref{spec}).

The nucleon vertex function in the non-covariant LC normalization
is defined by the pole behavior of the quark-diquark t-matrix in the variable $p_+ \,$:
\begin{eqnarray}
T(p) \rightarrow \frac{\Gamma_N(p) \, \overline{\Gamma}_N(p)}
{p_+ - \tilde{\epsilon}_p} \,\,\,\,\,\,\,\,\,\,\,\,\,\,\,{\rm as}\,\,\,
p_+ \rightarrow \tilde{\epsilon}_p      \label{lcn}\,,
\end{eqnarray}
where $\tilde{\epsilon}_p$ is the ``LC energy''. From this definition and Eq.(\ref{tn}) one obtains
\begin{eqnarray}
\Gamma_N(p)=\sqrt{- Z_N \, \frac{M_N}{p_{{\rm N}-}}} \,\, u_N(p_{\rm N})\,,  \label{gamma}
\end{eqnarray}
where $u_N$ is a free Dirac spinor for mass $M_N$ normalized as $\overline{u}_N u_N = 1$ and for an
on-shell nucleon we have
$p_{{\rm N}-}=(E_p-p_3)/\sqrt{2}$. The normalization factor $Z_N$ is given by
\footnote{Here, and in the following, the derivative w.r.t. $\fslash{p}$ of a Dirac matrix
valued function $a(p^2) + \fslash{p} b(p^2)$ is evaluated by using $p^2=\fslash{p}^2$ as
$2 \fslash{p} a'(p^2) + b(p^2) + 2 p^2 b'(p^2)$, where the prime denotes a derivative w.r.t. $p^2$.}
\begin{eqnarray}
Z_N=\left[\left(- \frac{\partial \Pi_{N0}(p)}{\partial \fslash{p}}\right)
_{\fslash{p}=M_N} \right]^{-1}\,.  \label{zn}
\end{eqnarray}
We note that in this normalization the vertex function satisfies the relation
\begin{eqnarray}
\overline{\Gamma}_N(p) \, \left(\frac{\partial \Pi_{N0}(p)}
{\partial p_+}\right) \, \Gamma_N(p) = 1. \label{nwf}
\end{eqnarray}
In the numerical calculations of this paper, we will approximate the quantity $\tau_0$ by
\begin{eqnarray}
\tau_0(q) \rightarrow 4i G_s - \frac{i\,g_D}{q^2-M_D^2}.   \label{pole}
\end{eqnarray}
Here the diquark mass $M_D$ is the pole of $\tau_0$ of Eq.(\ref{taus})
and the residue is given by
\begin{eqnarray}
g_D = 2 \left[ \left( - \frac{\partial \Pi_{s0}(q)}{\partial q^2}
\right)_{q^2=M_D^2}\right]^{-1}\,.
\label{gd}
\end{eqnarray}

\subsection{Evaluation of quark distributions via the moments}
The calculation of the distribution $f_{q/N0}$, which appears in
the convolution integral
(\ref{conv0}), in the quark-diquark model follows the lines explained
in Ref.\cite{MIN1} for the free nucleon. The difference, however,
is that here we are using the
proper time regularization scheme, which is defined in the
Euclidean formulation after
Wick rotation. In this case, the regularized LC momentum distribution cannot be
calculated directly from the Feynman diagrams of Fig.1,
since in the loop integrals one has to
fix the minus component
of the quark momentum to a real value, see Eq.(\ref{fey1}),
which is not possible in the Euclidean formulation.
We therefore consider first the moments
\begin{eqnarray}
A_n=\int_0^1 {\rm d}x\,x^{n-1}\,f_{q/N0}(x)\,, \label{moment}
\end{eqnarray}
and then derive 
an expression for the quark distributions which can easily be used
in Euclidean regularization schemes\footnote{A more general discussion
of this method can be found in Ref.\cite{NEW}.}.

In the following, we will outline the calculations for the ``quark diagram'' 
(first diagram of Fig. 1) and the ``diquark diagram''
(second diagram of Fig. 1) separately, leaving the details to
Appendix D~\footnote{Since all formulae
given below in this Subsection refer
to the case without the explicit
effect of the mean vector field, there is no difference between the
kinetic and canonical momenta and therefore we denote the
nucleon, quark, and diquark momenta as
simply $p,\,k,$ and $q$, respectively.}.
The contribution from the quark (diquark)
diagram will be denoted as $f_{q/N0}^{(Q)}(x)$ ($f_{q/N0}^{(D)}(x)$) and
the total isoscalar quark distribution in the nucleon is then given by the sum
\begin{eqnarray}
f_{q/N0}(x) = f_{q/N0}^{(Q)}(x) + f_{q/N0}^{(D)}(x)\,.  \label{tot}
\end{eqnarray}
Here, as in the previous subsection, the subindex ($0$) indicates
that the effect of the mean vector field is not included. 
The contribution of the quark diagram to the n-th
moment of $f_{q/N0}(x)$ is given
by~\footnote{See Eq.(3.7) of ref.\cite{MIN1}.
The nucleon vertex function (\ref{gamma}) in the
noncovariant LC normalization is
equal to the vertex function of ref.\cite{MIN1} multiplied by a factor
$1/\sqrt{2 p_{-}}$.}
\begin{eqnarray}
A_{n}^{(Q)} = - \overline{\Gamma}_N(p) \frac{\partial}{\partial p_+} \left(
\int \frac{{\rm d}^4 k}{(2\pi)^4}\, \left( \frac{k_-}{p_-}\right)^{n-1}\,S_0(k)\,\, \tau_0(p-k)
\right) \Gamma_N(p).  \label{anq}
\end{eqnarray}
From (\ref{nwf}) and (\ref{bubbn}) it follows
that the contribution of the quark diagram to the first
moment (quark number) is equal to $1$.

Because of the derivative in (\ref{anq}), only the pole term in the
expression (\ref{pole}) for $\tau_0$
contributes. Introducing a Feynman parameter ($\alpha$),
and observing that $g^{--}=0$, we obtain (see Appendix D)
\begin{eqnarray}
A_n^{(Q)} = -i g_D \, Z_N \, \frac{\partial}{\partial \fslash{p}}
\, \int \frac{{\rm d}^4 k}{(2\pi)^4}\, \int_0^1 {\rm d} \alpha\, \alpha^{n-1}
\frac{\left(\fslash{p} \alpha + M \right) + \frac{n-1}{2\alpha M_N}\left(k_0^2 - k_3^2\right)}
{\left(k^2 - A(\alpha,p^2)\right)^2}\,, \nonumber \\
\label{anq1}
\end{eqnarray}
where for the calculation of
the derivative it is understood that $p^2=\fslash{p}^2$ and at the end one
sets $\fslash{p}=M_N$. The quantity $A(\alpha,p^2)$ is defined as
\begin{eqnarray}
A(\alpha,p^2) = M^2(1-\alpha) + M_D^2 \alpha - p^2 \alpha (1-\alpha)\,.  \label{a}
\end{eqnarray}
From (\ref{zn}) we obtain the following form of the normalization factor $Z_N$:
\begin{eqnarray}
Z_N^{-1} = -i g_D \, \frac{\partial}{\partial \fslash{p}}
\, \int \frac{{\rm d}^4 k}{(2\pi)^4}\, \int_0^1 {\rm d} \alpha\,
\frac{\left(\fslash{p} \alpha + M \right)}{\left(k^2 - A(\alpha,p^2)\right)^2}\,,
\label{zn1}
\end{eqnarray}
and from Eqs. (\ref{anq1}), (\ref{zn1}) we see that $A_1^{(Q)}=1$.

If we perform a partial integration in $\alpha$ for the term $\propto n-1$ in 
Eq.(\ref{anq1}) and compare the resulting expression to the definition of
the moments (Eq.(\ref{moment})), we can directly read off the distribution function:
\begin{eqnarray}
f_{q/N0}^{(Q)}(x)= -i\, g_D \, Z_N\,\frac{\partial}{\partial \fslash{p}}
\, \int \frac{{\rm d}^4 k}{(2\pi)^4}\, \left( \fslash{p}x + M - \frac{k_0^2-k_3^2}{2 M_N}
\frac{\rm d}{{\rm d}x} \right) \frac{1}{\left(k^2 - A(x,p^2)\right)^2} \nonumber \\
\label{one}
\end{eqnarray} 
for $0<x<1$ and zero otherwise.
This expression can readily be used in Euclidean regularization schemes. 
Performing a Wick rotation and introducing the proper time regularization scheme 
(\ref{pt}), we finally obtain
\begin{eqnarray}
f_{q/N0}^{(Q)}(x) = \frac{g_D \, Z_N}{16 \pi^2} (1-x)
\int_{ 1/\Lambda_{\rm UV}^2}^{1/\Lambda_{\rm IR}^2} \frac{{\rm d}\tau}{\tau} e^{-\tau A(x,M_N^2)}
\left[1+  \tau x \left(\left(M_N+M\right)^2-M_D^2\right) \right]\,. \nonumber \\
\label{fxq}
\end{eqnarray}

The contribution of the diquark diagram of Fig.1
to the distribution $f_{q/N0}$ can be expressed as
a convolution integral, involving the quark
distribution in a virtual diquark ($f_{q/D0}$)
and the diquark distribution in the nucleon ($f_{D/N0}$).
For simplicity we use here the on-shell
approximation for the diquark distribution, that is,
we assume that $f_{D/N0}$
has a sharp peak at the diquark virtuality
$q_0^2=M_D^2$, so that $f_{q/D0}$ can
be replaced by its on-shell value
and taken out of the integral over the virtuality.
Then we use the relation \cite{MIN1}
${\displaystyle \int {\rm d}q_0^2\,f_{D/N0}(y,q_0^2)=f_{q/N0}^{(Q)}(1-y)}$,
where $f_{q/N0}^{(Q)}$
was evaluated above, see (\ref{fxq}).
The contribution of the diquark diagram is then given by
\begin{eqnarray}
f_{q/N0}^{(D)}(x)=\int_0^1 {\rm d}y \int_0^1 {\rm d}z\,
\delta(x-yz) f_{q/D0}(z)\,f_{q/N0}^{(Q)}(1-y)\,.
\label{fd}
\end{eqnarray}
Here $f_{q/N0}^{(Q)}$ is given by Eq.(\ref{fxq})
and we anticipated the fact that $f_{q/D0}(z)$
has support for $0<z<1$. To evaluate $f_{q/D0}(z)$, we use the moments
(see Eq.(3.10) of Ref.\cite{MIN1})
\begin{eqnarray}
A_n^{(q/D)}=- 6i\,g_D \, \frac{\partial}{\partial q^2}
\int \frac{{\rm d}^4 k}{(2\pi)^4}
\left(\frac{k_-}{q_-}\right)^{n-1}\,{\rm Tr}_D \left(\gamma_5 S_0(k)
\gamma_5 S_0(k-q)\right) \,,
\label{anqd}
\end{eqnarray}
where one has to set $q^2=M_D^2$ at the end.

Before giving the explicit expressions, let us confirm the number
and momentum sum rules.
From the form of $g_D$ given in (\ref{gd}), it follows
that $A_1^{(q/D)}=2$,
and because we have already confirmed that the $n=1$ moment of
$f_{q/N0}^{(Q)}$ is equal to $1$,
the convolution (\ref{fd}) shows that the diquark diagram gives a
contribution equal to $2$ to the number sum rule. The number sum rule value for the total distribution
(\ref{tot}) is therefore
\begin{eqnarray}
\int_0^1  {\rm d}x\, f_{q/N0}(x) = 3 \,.   \label{totns}
\end{eqnarray}
For the momentum sum rule we obtain
\begin{eqnarray}
\int_0^1  {\rm d}x\,x \, f_{q/N0}(x) &=& \int_0^1  {\rm d}x\,x \, f_{q/N0}^{(Q)}(x) \nonumber \\
&+& \left(\int_0^1  {\rm d}y\,y \, f_{q/N0}^{(Q)}(1-y)\right) \cdot
\left(\int_0^1  {\rm d}z\,z \, f_{q/D0}(z)\right)\,. \nonumber \\
\label{msum1}
\end{eqnarray}
Using the fact that $f_{q/D0}(z)$ is symmetric around
${\displaystyle z=\frac{1}{2}}$ and that its $n=1$ moment
is equal to $2$, we see that
${\displaystyle \int_0^1  {\rm d}z\,z \, f_{q/D0}(z) = 1}$. The momentum
sum rule (\ref{msum1}) then becomes
\begin{eqnarray}
\int_0^1
{\rm d}x\,x \, f_{q/N0}(x) = \int_0^1  {\rm d}x\,x \, f_{q/N0}^{(Q)}(x)
+ \int_0^1  {\rm d}x\,(1-x) \, f_{q/N0}^{(Q)}(x) = 1\,. \label{msum2}
\end{eqnarray}

We now continue with the
evaluation of (\ref{anqd}) and the distribution $f_{q/D0}(x)$.
By using a Feynman parameter ($\alpha$) and observing that $g^{--}=0$ we obtain
(see Appendix D)
\begin{eqnarray}
A_n^{(q/D)} &=& - 24 i \, g_D \, \frac{\partial}{\partial q^2} \int \frac{{\rm d}^4 k}{(2\pi)^4}
\int_0^1 {\rm d}\alpha \, \alpha^{n-1}  \nonumber \\
&\times&   
\frac{M^2+q^2 \alpha (1-\alpha) -k^2 + \frac{n-1}{2\alpha}(1-2 \alpha) (k_0^2 - k_3^2)}
{\left(k^2 + q^2 \alpha (1-\alpha) - M^2 \right)^2}\,.  \label{anqd1}
\end{eqnarray}
If we perform a partial integration in $\alpha$ for the term $\propto n-1$ in Eq.(\ref{anqd1})
and compare to the definition of the moments (Eq.(\ref{moment}), we obtain 
\begin{eqnarray}
f_{q/D0}(x) &=& - 24 i \, g_D \, \frac{\partial}{\partial q^2} \int \frac{{\rm d}^4 k}{(2\pi)^4}
\left(M^2+q^2 \alpha (1-\alpha) -k^2 - \frac{k_0^2-k_3^2}{2} \frac{\rm d}{{\rm d}x}
\left(1-2x\right)\right) \nonumber \\
&\times& \frac{1}{\left(k^2+q^2 x (1-x) - M^2\right)^2}\,, \label{onep}
\end{eqnarray}
where the derivative ${\rm d}/{\rm d}x$ acts on the whole function on its r.h.s. 
Performing a Wick rotation and introducing the proper time regularization (\ref{pt}), we
finally obtain
\begin{eqnarray}
f_{q/D0}(x) = \frac{3 g_D}{4\pi^2}
\int_{ 1/\Lambda_{\rm UV}^2}^{1/\Lambda_{\rm IR}^2} \frac{{\rm d}\tau}{\tau}\,
e^{-\tau \left(M^2-M_D^2 x(1-x)\right)} \left(1+M_D^2 \tau x (1-x)\right) \,. \nonumber \\
\label{fxd}
\end{eqnarray}
The results (\ref{fxq}) and (\ref{fxd}) are then inserted into the convolution formula (\ref{fd}) to give the
contribution of the diquark diagram to the quark distribution function.

\subsection{Model for nuclear matter}
The EOS of NM in the mean field approximation to the NJL model
can be derived in a formal way
\cite{QM} from the quark Lagrangian (\ref{lag1}) by using hadronization
techniques. Equivalently, but more intuitively, one can use
the ``hybrid model
assumption'' that the expectation value of
any local quark operator, like the quark Hamiltonian density, can be split into
the expectation value in the valence quark
vacuum $|0\rangle=|\rho=0\rangle$ and an average over the
nucleon Fermi sea consisting of correlated valence quarks.
The resulting energy
density of isospin symmetric NM at rest has the form \cite{BT,QM}
\begin{eqnarray}
{\cal E}={\cal E}_V - \frac{V_{0}^2}{4G_{\omega}} + 4 \int
\frac{d^3 p}{(2 \pi)^3}\, \Theta\left(p_F-|{\bold p}|\right)\, \epsilon_p,
\label{en}
\end{eqnarray}
where $\epsilon_p$ is
given by (\ref{spec}) and the vacuum contribution (quark loop) is
\begin{eqnarray}
{\cal E}_V=12i \int \frac{{\rm d}^4 k}{(2\pi)^4} \,
{\rm ln}\, \, \frac{k^2-M^2+i\epsilon}{k^2-M_0^2+i\epsilon}
+\frac{(M-m)^2}{4 G_{\pi}} -\frac{(M_0-m)^2}{4 G_{\pi}}\,.
\label{env}
\end{eqnarray}
The condition $\partial {\cal E} / \partial V_0 = 0$ leads to
\begin{equation}
V_0=6 \, G_{\omega}\, \rho\,,  \label{solv}
\end{equation}
and we can eliminate the vector field in (\ref{en})
in favor of the baryon density. The resulting expression has then
to be minimized with respect to the constituent quark
mass $M$ for fixed density. For zero density this
condition becomes identical to the
familiar gap equation of the NJL model, and for finite
density the nonlinear $M$-dependence of the
nucleon mass $M_N$, which is obtained from the
pole of $T_0(p)$ (see Eq.(\ref{tn})), is
essential to obtain saturation of the binding energy per nucleon~\cite{BT}.

\section{Results and discussions}
\setcounter{equation}{0}
In this Section we present our numerical results for the quark LC momentum
distributions, the structure functions and the EMC effect.
The parameters used in
the calculation are the same as in Refs.~\cite{BT,QM}.
The IR cut-off $\Lambda_{\rm IR}$ is fixed at 200 MeV, while the values of
$m$, $G_{\pi}$ and $\Lambda_{\rm UV}$ are
determined so as to reproduce $M_0=400$ MeV by the
gap equation in the vacuum, $m_{\pi}=140$ MeV from
the pole of the t-matrix in the pionic
$q \overline{q}$ channel and $f_{\pi}=93$ MeV from
the one quark loop diagram for pion decay. The resulting
values are $m=16.9$ MeV, $G_{\pi}=19.60$ GeV$^{-2}$,
and $\Lambda_{\rm UV}=638.5$ MeV.
The coupling constant $G_s$ of (\ref{lags}) is
determined so as to reproduce $M_{N0}=940$ MeV
as the pole of the quark-diquark t-matrix
at zero density (see Eq.~(\ref{tn})),
which gives the ratio $r_s=G_s/G_{\pi}=0.508$.
Finally, the coupling constant $G_{\omega}$ is
determined so that the curve for the NM
binding energy per nucleon $(E_B/A$) as a function of the density
passes through the empirical saturation
point ($\rho, E_B/A)=(0.16$ fm$^{-3},\,15$ MeV),
which gives the ratio $r_{\omega}=G_{\omega}/G_{\pi}=0.37$.
{}From the pole of the t-matrix in the vector $q \overline{q}$ channel,
this value gives an
$\omega$ meson mass in vacuum\cite{QM} of $827$ MeV,
which is larger than the experimental value
($783$ MeV), but is nevertheless reasonable.

We recall from Ref.\cite{BT} that, in this simple NJL model, $G_{\omega}$
is the only free parameter for the NM calculation.
Therefore, although the binding energy curve
passes through the empirical saturation point,
its minimum is at a different point,
($\rho, E_B/A)=(0.22$ fm$^{-3},\,17.3$ MeV), see Fig. 6 of Ref.\cite{BT}. As we
discussed in Sect.2, in the derivation of
Eq.~(\ref{fqa1}) for the quark distribution in NM
we assumed that $\overline{M}_N=\epsilon_F$, which holds only at the minimum of the binding
energy curve.
Therefore, although our calculated saturation density $\rho=0.22$ fm$^{-3}$ is
larger than the empirical one,
we have to use this density in our numerical calculations
for consistency~\footnote{However,
in order to indicate the density dependence of our results, we
also show in Appendix E the case $\rho=0.16$ fm$^{-3}$, simply by using
Eq.(\ref{fqa1}) for this value of the density.}.

In Table 1 we list the effective quark,
diquark and nucleon masses, the vector field acting on
the quark ($V_0$), the nucleon Fermi energy
and the mass per nucleon $\overline{M}_N=M_{N0}-E_B/A$
at $\rho=0$, $\rho=0.16$ fm$^{-3}$, and $\rho=0.22$ fm$^{-3}$.
We note that for $\rho=0.22$ fm$^{-3}$ the sum $M+M_D$ is
smaller than $M_N$, which is
possible because of the absence of
unphysical thresholds in our regularization procedure.
\begin{table}[h]
\begin{center}
\begin{tabular}{|c|c|c|c|}
\hline
  &  $\rho=0$     &  $\rho=0.16$fm$^{-3}$  &  $\rho=0.22$fm$^{-3}$  \\  \hline
$M$               &  400  &  308  &  276  \\
$M_D$             &  576  &  413  &  355  \\
$M_N$             &  940  &  707  &  634  \\
$V_0$             &  0    &   53  &   75  \\
$\epsilon_F$      &  940  &  914  &  923  \\
$\overline{M}_N$  &  940  &  925  &  923  \\  \hline
\end{tabular}
\end{center}
\caption{Effective masses $M$, $M_D$ and $M_N$ for the quark, the diquark and the nucleon,
the vector field for quarks ($V_0$), the nucleon Fermi energy ($\epsilon_F$), and the
mass per nucleon $\overline{M}_N=M_{N0}-E_B/A$ (all in MeV)
at $\rho=0$, $\rho=0.16$ fm$^{-3}$ and
$\rho=0.22$ fm$^{-3}$.  The saturation density in the
present calculation is $\rho=0.22 $fm$^{-3}$.
The values of the quark-diquark coupling
constant $g_D$ (see Eq.(\ref{pole})) are
18.23, 16.16, and 15.64 for $\rho=0$,
$\rho=0.16$ fm$^{-3}$, and $\rho=0.22$ fm$^{-3}$, respectively.}
\end{table}

We first show our results for the valence up and down
quark distributions in the
free proton in Figs. 3 and 4. They are given by
\begin{eqnarray}
u_v(x) &=& f_{q/N0}^{(Q)}(x) + \frac{1}{2}\, f_{q/N0}^{(D)}(x)\,,
\label{uv} \\
d_v(x) &=& \frac{1}{2} \, f_{q/N0}^{(D)}(x)\,,  \label{dv}
\end{eqnarray}
where the contributions from the quark
diagram (Eq.~(\ref{fxq})) and the diquark diagram
(Eq.~(\ref{fd})) refer to the case $\rho=0$ of Table 1.

\begin{figure}[h]
\begin{center}
\epsfig{file=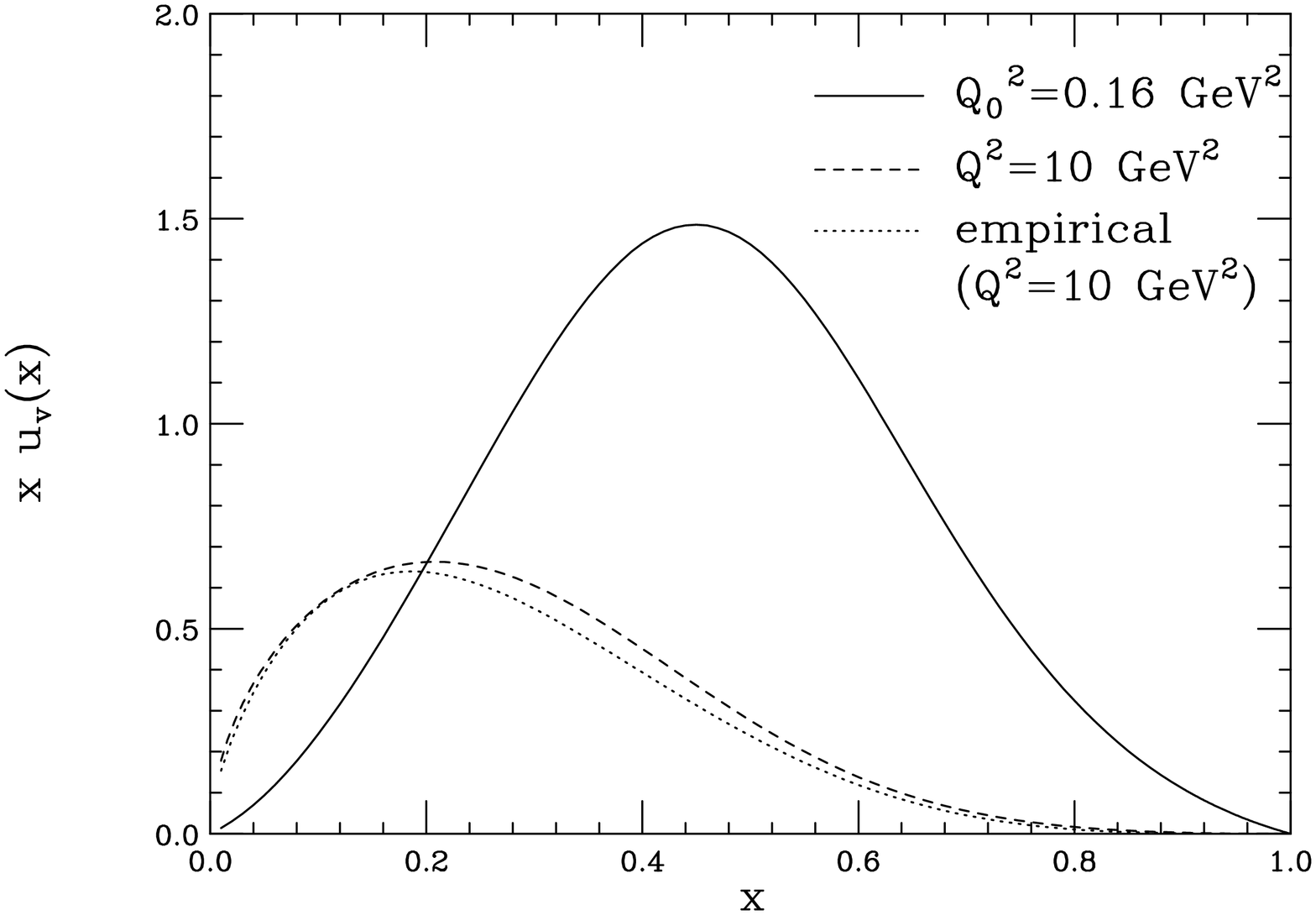,angle=90,width=11cm}
\caption{The valence up quark LC momentum distribution
in the free nucleon, multiplied by
$x$. The solid line shows the NJL model
result ($Q_0^2=0.16$ GeV$^2$), the dashed
line shows the result after the evolution up to $Q^2=10$ GeV$^2$, and the dotted
line shows the empirical parametrization of Ref.\cite{MSR}.}
\end{center}
\end{figure}

\begin{figure}[h]
\begin{center}
\epsfig{file=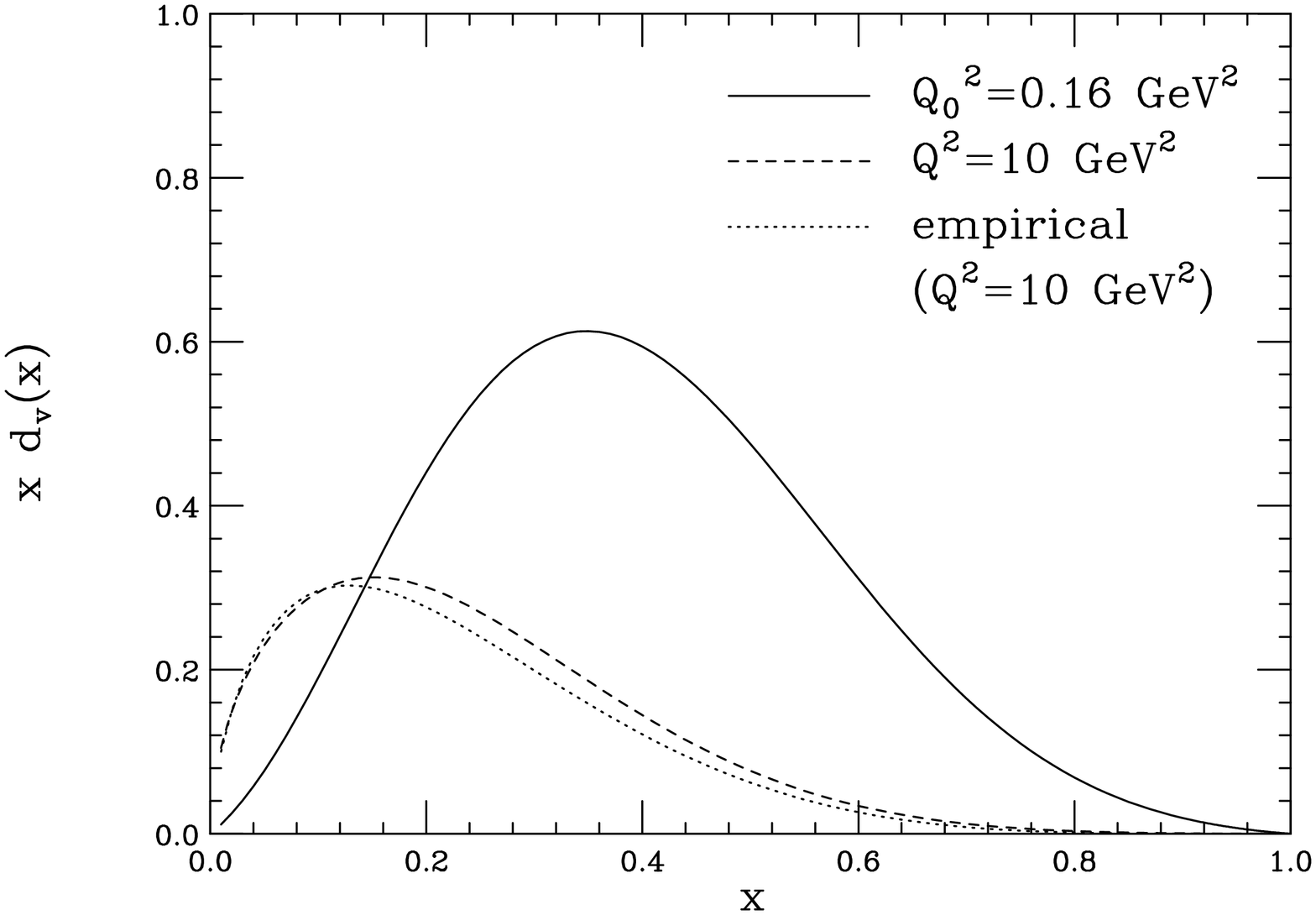,angle=90,width=11cm}
\caption{Same as Fig. 3 for the valence down quark.}
\end{center}
\end{figure}

The solid lines show our NJL model results. We see
clearly that $d_v(x)$ is softer than $u_v(x)$ --
that is, the peak position of $d_v(x)$ is at smaller values of $x$.
This difference comes from
the diquark correlations \cite{SDC,MIN2}, or,
in other words, from the difference between the ``hard'' distribution
$f_{q/N0}^{(Q)}$ and the ``soft''
one $f_{q/N0}^{(D)}$. The total LC momentum fraction carried by the
$u$ quarks is $71\%$ and that of the $d$ quark is $29\%$.

In order to make contact to the measured structure functions, we associate the
distributions calculated in the NJL model
with some low energy scale $Q_0$, where a description in terms of
quark degrees of freedom alone is expected to be valid.
We then perform the evolution
according to the Dokshitzer-Gribov-Lipatov-Altarelli-Parisi (DGLAP) equations
\cite{DGLAP} up to the value of $Q$ where experimental data
for DIS and empirical parametrizations of parton distributions are avaliable
\footnote{We use the computer code of Ref.\cite{MK} with
$N_f=3$, $\Lambda_{\rm QCD}=250$ MeV
in the $\overline{\rm MS}$ renormalization and
{}factorization scheme in the next to
leading order to perform the $Q^2$ evolution.}.
Here we fix $Q_0=400$ MeV as in our earlier calculations \cite{MIN1,MIN2},
and perform the evolution up to $Q^2 = 10$ GeV$^2$.
The resulting valence quark distributions
are shown by the dashed lines in Figs. 3 and 4 and they are compared with
the empirical distributions of Ref.\cite{MSR} shown by the dotted lines.
We see that our calculation
reproduces the main features of the
empirical distributions. The calculated total LC momentum fractions
at $Q^2=10$ GeV$^2$ are $29\%$ and $12\%$ for the valence $u$ quarks
and the valence $d$ quark, respectively,
while the rest is carried by the gluons and sea quarks which are
generated by the $Q^2$ evolution.

We now discuss the modifications of the quark LC momentum
distribution at finite density.
For this purpose, we first replace the free quark, diquark and
nucleon masses by the effective
ones given in the last column of Table 1, which gives the
distribution $f_{q/N0}$. Then we calculate the
convolution with the nucleon LC momentum distribution
according to Eq.(\ref{conv0}) to obtain $f_{q/A0}$. Finally we include
the effect of the vector potential
through the scale transformation
(\ref{fqa1}) to get the total result, $f_{q/A}$.
The results are shown in Fig. 5.
The dotted line shows isoscalar valence quark distribution for the
free nucleon and the dashed line shows the result when all masses are replaced
by the effective ones. We see that the distribution becomes somewhat
stiffer  -- i.e., it has more support for large
momentum fractions. This can be understood
as follows: For large $x$ the spectator quark,
which corresponds to the quark diagram of Fig. 1,
gives the most important contribution. From
the spectral representation (\ref{spq}) it
follows that the typical LC momentum fraction carried by the spectator quark
is $1-M_D/M_N$ -- in agreement with Ref.~\cite{SDC}.
Since in our calculation the ratio $M_D/M_N$ decreases as the density
increases (see Table 1), the fraction carried by the spectator quark increases
\footnote{We wish to point out that, in spite of the fact that the LC momentum
distribution becomes stiffer, the baryon radius of the
nucleon increases in the medium~\cite{TOD03},
which shows that there is no direct connection between the change of the
LC momentum distribution and the baryon radius.}.

\begin{figure}[h]
\begin{center}
\epsfig{file=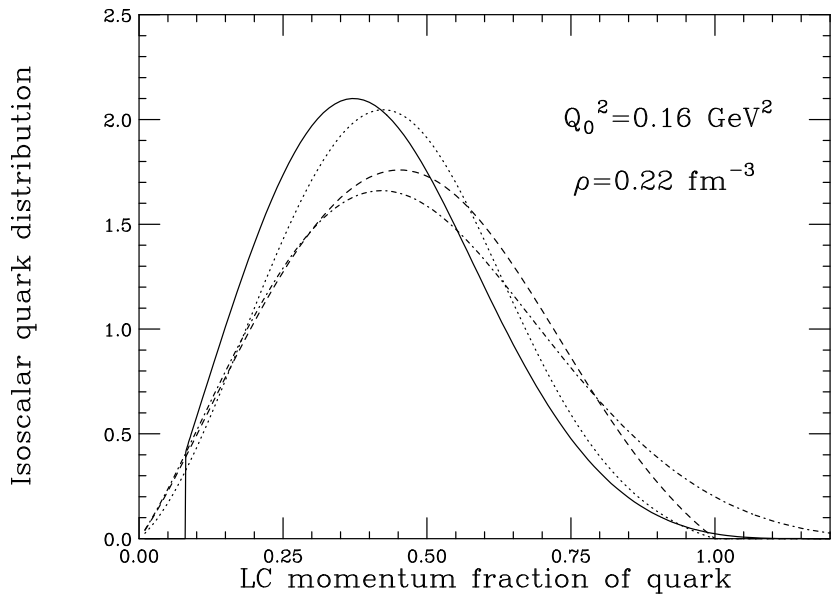,angle=90,width=11cm}
\caption{Sum of the valence up and down quark LC
momentum distributions at the low energy
scale ($Q_0^2=0.16$ GeV$^2$), multiplied by the LC momentum fraction. 
The dotted line shows the distribution in the free
nucleon, the dashed line is obtained
by replacing the free masses by the effective
ones, the dot-dashed line shows the result
of the convolution (\ref{conv0}), and the
solid line is the final result
obtained from the scale transformation (\ref{fqa1}).
The LC momentum fraction on the horizontal axis
generically refers to the proper variable of each
distribution, i.e., $x$ for the dotted and
dashed lines, and $x_A$ for the dot-dashed and
solid lines.}
\end{center}
\end{figure}

The result of the convolution (\ref{conv0})
with the nucleon momentum distribution
is shown by the dot-dashed line in Fig. 5.
Because the nucleon momentum distribution $f_{N/A0}(y)$ has a finite
width and is symmetric around $y=1$
(see Eq.(\ref{fn0})), the result of the convolution shows the familiar
effects of Fermi motion. The support extends beyond $x=1$ and,
relative to the dashed line, there is a depletion
for intermediate values of $x$. However, we
have to note two points here. First, since at this stage the
effect of the vector field is not yet included,
the convolution (\ref{conv0}) involves the nucleon momentum distribution
(\ref{fn0}) {\em without} the explicit effect of the
vector potential, and has an appreciably
larger width than the ``full'' distribution (see (\ref{rya}) and (\ref{rya0})).
Therefore, the effect of the Fermi motion, which is seen by comparing
the dashed and dot-dashed lines in Fig. 5,
is more pronounced than in the calculations of
Ref.\cite{MIL}, where the convolution of the
quark distribution in a free nucleon with
the {\em full} $f_{N/A}$ was considered. Second,
Fig. 5 clearly shows that the EMC effect
cannot be explained by taking into account only
the effects of the scalar potential
and Fermi motion: in comparison with the dotted line, the dot-dashed
line shows an enhancement in the valence quark
region ($x\simeq 0.7$) instead of a quenching.

We now come to the discussion of the effect of the mean vector
field on the quark distribution function, as expressed by Eq.(\ref{fqa1})
and shown by the solid line in Fig. 5. If we compare the solid line
with the dot-dashed line, we see that the
effect of the vector field is to squeeze the quark distribution
both from the small and large
$x$ side, leading to a reduction of the width, a quenching in the valence quark
region and an enhancement in the region of smaller $x$.
To understand the origin of this
effect, we recall the key points which led to the scaling
relation (\ref{fqa1}):
The presence of the vector field leads to a shift of
the quark momentum according to $k^{\mu}=k_{\rm Q}^{\mu} + V^{\mu}$,
see Eq.(\ref{kinh}) for $H=1$, and of the nucleon chemical
potential (mass per nucleon) according
to $\epsilon_F=E_F+3 V_0$, see Eq.(\ref{hh}).
Then the ratio of the quark LC momentum
to the chemical potential without the effect of the vector field
(${\displaystyle x_A'= \frac{k_{{\rm Q}-}}{E_F/\sqrt{2}}}$)
is related to the one including the
effect of the vector field
(${\displaystyle x_A=\frac{k_-}{\epsilon_F/\sqrt{2}}}$) by
\begin{eqnarray}
x_A' = \frac{k_- - V_-}{E_F/\sqrt{2}} =
x_A \frac{\epsilon_F}{E_F}-\frac{V_0}{E_F},
\label{int}
\end{eqnarray}
which is just the rescaling of the variables
expressed in Eq.(\ref{fqa1}). Using (\ref{hh}), this
linear relationship between $x_A'$ and $x_A$ can also be expressed as
\begin{eqnarray}
x_A' = x_A\left(1+3 v\right) - v\,,
\label{lin}
\end{eqnarray}
where $v=V_0/E_F$. From Eq.(\ref{lin}) we see that for
${\displaystyle x_A=\frac{1}{3}}$ we also have
${\displaystyle x_A'=\frac{1}{3}}$
and for ${\displaystyle x_A>\frac{1}{3}}$ we have $x_A'>x_A$, while for
${\displaystyle x_A<\frac{1}{3}}$ we have $x_A'<x_A$.
Thus, if $f_{q/A0}(x_A)$ has the
shape of a typical valence quark distribution with a peak around
${\displaystyle x_A=\frac{1}{3}}$, the distribution
${\displaystyle f_{q/A0}(x_A'=x_A \frac{\epsilon_F}{E_F}-\frac{V_0}{E_F})}$,
viewed as a function of $x_A$, will be squeezed from both sides
($x_A<1/3$ and $x_A>1/3$),
with the amount of squeezing being larger for large $x_A$.
According to (\ref{fqa1}),
this distribution is then multiplied by an overall factor
${\displaystyle \frac{\epsilon_F}{E_F}}$ so as to satisfy the number
and momentum sum rules.
This mechanism, which is clearly seen by comparing the dot-dashed and the solid curves in Fig. 5
\footnote{The jump at small $x_A$ shown by the solid line in Fig.5 can also be understood from
the rescaling relation Eq.(\ref{fqa1}): The original distribution $f_{q/A0}(x_A')$ is nonzero at 
$x_A'=0$ but zero for $x_A'<0$, and therefore the shifted distribution $f_{q/A}(x_A)$ is discontinuous
at $x_A=V_0/\epsilon_F$. However, in this region one has to consider also the effect of the sea quark
distributions, which are not included here.}, is largely responsible for the explanation of the 
EMC effect in a mean field description of NM, where the vector field couples to the quarks inside the nucleons.

If we finally compare the
dotted and solid curves of Fig. 5,
we observe an enhancement for momentum fractions between $0.2$
and $0.4$, and a depletion for larger $x$,
besides the enhancement for $x\rightarrow 1$ arising
from the Fermi motion. These features are consistent with the EMC effect.

We next show in Fig. 6 our results for the
nuclear structure function per nucleon, calculated for isospin
symmetric NM,
\begin{eqnarray}
F_{2A}(x_A) = \frac{5}{18} \, x_A \, f_{q/A}(x_A)\,,  \label{f2a}
\end{eqnarray}
in comparison to the isoscalar structure function of a free nucleon
\begin{eqnarray}
F_{2N}(x) = \frac{1}{2} \left(F_{2p}(x) + F_{2n}(x)\right) =
\frac{5}{18} \, x \,
f_{q/N0}(x)\,.  \label{f2n}
\end{eqnarray}

\begin{figure}[h]
\begin{center}
\epsfig{file=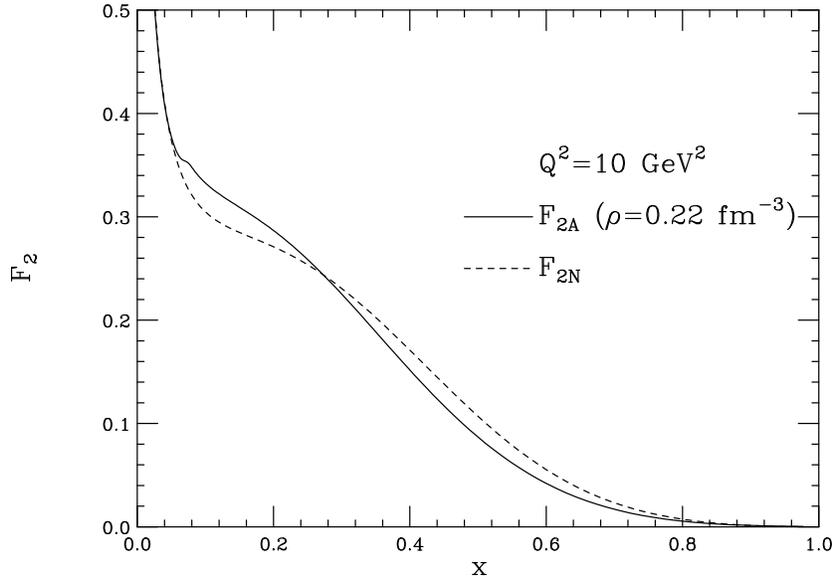,angle=90,width=11cm}
\caption{The solid line shows the structure function per nucleon in symmetric NM
given by Eq.(\ref{f2a}). It is plotted as a function of $x$ by using the
relation ${\displaystyle x_A=\frac{M_{N0}}{{\overline M}_N}\, x}$.
The dashed line shows the isoscalar free nucleon
structure function of Eq.~(\ref{f2n}).}
\end{center}
\end{figure}

The curves of Fig. 6 were obtained by
evolving the quark distributions shown by the
solid and dotted lines in Fig. 5 up to $Q^2=10$ GeV$^{2}$. In order to plot the
nuclear structure function (\ref{f2a})
as a function of $x$, we used the relation
${\displaystyle x_A=\frac{M_{N0}}{\epsilon_F}\, x = 1.02\, x}$,
see Eqs.(\ref{bja}), (\ref{hh})
and Table 1. Although this ``binding effect'' is on the level of a few percent,
it gives a non-negligible contribution at large $x$.
Fig. 6 clearly shows the suppression of the structure function at large
$x$ and the enhancement at smaller $x$, which we discussed above for the
distribution function at the low energy scale.

From the two curves of Fig. 6, we obtain the ratio between the
nuclear and the nucleon structure functions
\begin{eqnarray}
R_{A/N}(x)=\frac{F_{2A}(x_A)}{F_{2N}(x)}\,,  \label{emcn}
\end{eqnarray}
which is shown in Fig. 7. The data points in this figure
are the extrapolations of nuclear DIS data to the NM case \cite{SD}
\footnote{The fit in Ref.\cite{SD} uses data over a wide range of $Q^2$, assuming
that the EMC ratio is fairly independent of $Q^2$. Although Fig. 7 refers to
$Q^2=10$ GeV$^2$, we have confirmed that our results are independent of
$Q^2$, as long as $Q^2 \agt 4$ GeV$^2$.}.

\begin{figure}[h]
\begin{center}
\epsfig{file=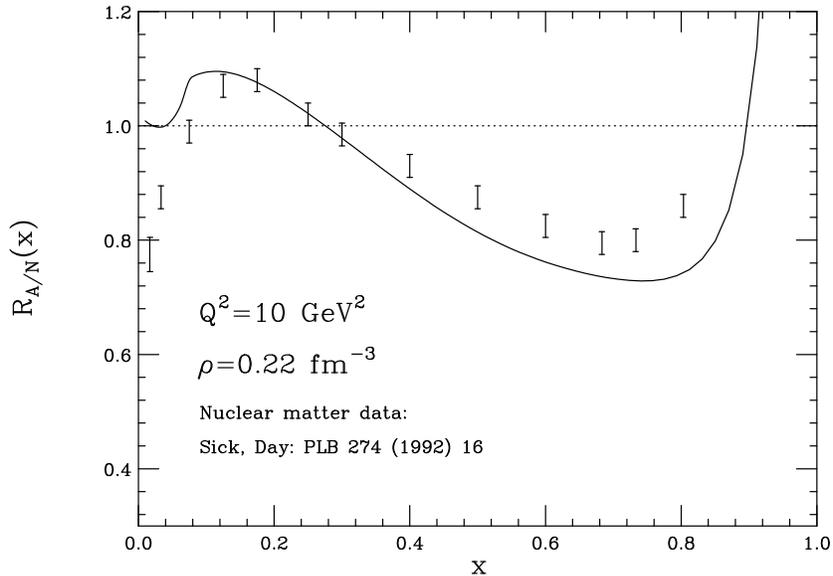,angle=90,width=11cm}
\caption{The ratio (\ref{emcn}) of the nuclear
to the nucleon structure function. The data points
are the extrapolations of nuclear DIS data to the NM case \cite{SD}.}
\end{center}
\end{figure}

We see that the calculation can reproduce the main features of the EMC effect,
namely the suppression at large $x$ and the enhancement at smaller $x$.
[The fact that in our calculation the saturation density is too large
($\rho=0.22$ fm$^{-3}$) leads to an EMC effect that is slightly too strong at large $x$.
This is clear if we compare to the results discussed in Appendix E for
$\rho=0.16 $ fm$^{-3}$ -- see Fig. 11.] 

\section{Summary and conclusions}
In this paper we have discussed the modifications of the structure function
of a nucleon bound in the nuclear medium. For this purpose, we used an
effective chiral quark theory which can account for the quark substructure
of a single nucleon, as well as for the saturation properties of the NM binding
energy. Our description of NM was carried out in the mean field approximation,
which is characterized by mean scalar and vector fields coupling to
the quarks inside the nucleons. Our aim was to assess the effects which arise
from the quark structure of the bound nucleon.

It is well known that the effects of the mean scalar and vector fields on the
level of nucleons, including the Fermi motion, cannot explain the EMC effect.
We have investigated these medium modifications
at the level of the quark structure
and our findings can be summarized as follows. The mean scalar
field can simply be incorporated into the effective masses and
together with the
Fermi motion of nucleons it cannot explain the observed
reduction of the EMC ratio in the intermediate $x$ region. The vector field, on the
other hand, influences the form of the quark LC momentum
distributions directly -- on top of
its indirect influence through the EOS of the system. This direct
modification of the quark distribution by the mean vector field is expressed
by Eq.(\ref{fqa1}). As our discussion in Sect.4 showed,
it plays a key role in explaining the EMC effect,
in the framework of a mean field description of NM in terms of quarks.
We note that the relation (\ref{fqa1}) includes the combined effect
of the mean vector field on the
quark momentum distribution in the nucleon and on the
nucleon momentum distribution in NM. While the effect of the mean
vector field on the nucleon momentum distribution alone tends to cancel the
effect of the scalar field \cite{MIL}, its combined effect is appreciable
and works in a direction which is consistent
with the EMC data. It is therefore crucial to
include the quark substructure of the nucleon for the explanation of the
EMC effect.

Our findings are reminiscent of the discussions in Ref.~\cite{BT} (and,
of course Refs.~\cite{GUI}), which
emphasized the importance of the fact that the mean {\em scalar} field
couples to the quarks in the nucleon, instead of to the nucleon as a whole,
for the stability of the NM ground state in chiral effective quark theories.
Concerning the mean {\em vector} field, on the other hand,
the quark substructure
of the nucleon is not essential as long as we consider only the bulk properties
of the system. It {\em is} essential, however, for the structure functions and
the EMC effect, as we have shown in this paper.
The further study of the influence of
the quark substructure on medium modifications
of nucleon properties along these lines
is an important aspect of what has come to be called quark nuclear physics.

\vspace{1cm}

{\sc Acknowledgment} \\

This work was supported by the Grant in Aid for Scientific
Research of the Japanese Ministry of
Education, Culture, Sports, Science and Technology, Project No. C2-13640298,
the Australian Research Council and The University of Adelaide.
The authors wish to thank
K. Saito, F. Steffens, K. Tsushima and A. Williams for helpful discussions.

\newpage

\appendix

{\LARGE Appendices}

\section{Verification of Eq.(\ref{relation}) from Feynman diagrams}
\setcounter{equation}{0}
In this Appendix we show that the result of an explicit evaluation of the Feynman diagrams shown
in Fig.1 satisfies the general relation (\ref{relation}).

First we consider the quark diagram of Fig.1, which gives the following contribution to the
distribution function (see Eq.(3.7) of Ref.\cite{MIN1}):
\begin{eqnarray}
f_{q/N}^{(Q)}(x) = {\overline \Gamma}_N(p) \left(\frac{\partial}{\partial p_+} \Pi_N(x,p)\right)\Gamma_N(p)
\label{fqn1}
\end{eqnarray}
with
\begin{eqnarray}
\Pi_N(x,p) = - \int \frac{{\rm d}^4k}{(2\pi)^4} \delta\left(x-\frac{k_-}{p_-}\right) S(k) \tau(p-k)\,. \label{pinz}
\end{eqnarray}
Here we make use of (see (\ref{qf}) and (\ref{taus}))
\begin{eqnarray}
S(k)=S_0(k_{\rm Q}) \,\,; \,\,\,\,\,\,\,\,\,\,
\tau(p-k)=\tau_0\left((p-k)_{\rm D}\right) = \tau_0(p_{\rm N} - k_{\rm Q})\,,  \label{pro}
\end{eqnarray}
where the kinetic momenta are defined
as in (\ref{kinh}) for H=Q, D and N, corresponding to the quark
numbers $H=1,\,2,\,3$. If we make a shift $k \rightarrow k + V$ in
(\ref{pinz}), the $\delta$ function becomes
\begin{eqnarray}
\delta\left(x-\frac{k_- +V_-}{p_-}\right) = \frac{p_-}{p_{{\rm N}-}}\,
\delta\left(x' - \frac{k_-}{p_{{\rm N}-}}\right)\,, \label{delta}
\end{eqnarray}
where ${\displaystyle x'=\frac{p_-}{p_{{\rm N}-}}x - \frac{V_-}{p_{{\rm N}-}}}$.
By noting that the on-shell value of $p_{\rm N}^0$ is
$E_p$, see Eq.(\ref{spec}), and that the
nucleon spinor $\Gamma_N(p)$ of Eq.(\ref{gamma}) differs from the one
{}for a free nucleon (quark-diquark bound state)
only by the replacement of the free nucleon mass by the
effective one, we immediately obtain the relation
\begin{eqnarray}
f_{q/N}^{(Q)}(x) = \frac{p_-}{p_{{\rm N}-}}\,f_{q/N0}^{(Q)}(x'=\frac{p_-}{p_{{\rm N}-}}x - \frac{V_-}{p_{{\rm N}-}})
\,, \label{qq}
\end{eqnarray}
where the distribution $f_{q/N0}^{(Q)}(x')$ differs from the expression for a free nucleon only by the
replacement of the free masses by the effective ones.

In the on-shell approximation for the diquark,
to which our actual calculations of Subsect.3.2 refer,
the contribution of the diquark diagram of Fig.1 to the quark distribution function
has the form \cite{MIN1}
\begin{eqnarray}
f_{q/N}^{(D)}(x) = \int {\rm d}y \, \int {\rm d}z \, \delta(x-yz) f_{q/D}(z) \, f_{D/N}(y)\,, \label{convd}
\end{eqnarray}
where
\begin{eqnarray}
f_{D/N}(y) = f^{(Q)}_{q/N}(1-y) = \frac{p_-}{p_{{\rm N}-}} f_{D/N0}\left(y' = \frac{p_-}{p_{{\rm N}-}}y -
\frac{2 V_-}{p_{{\rm N}-}}\right)\,.
\label{fdn}
\end{eqnarray}
To derive the second equality in (\ref{fdn}), we used the result (\ref{qq}), and
$f_{D/N0}(y')=f^{(Q)}_{q/N}(1-y')$.
The quark distribution in an on-shell diquark is given by (cf. Eq.~(3.19)
of ref.\cite{MIN1})
\begin{eqnarray}
f_{q/D}(z)= \frac{2}{\partial \Pi_s(q)/\partial q_+}\,\frac{\partial \Pi_s(z,q)}{\partial q_+}
\label{ffd}
\end{eqnarray}
at the on-shell value $q_{\rm D}^2=M_D^2$.
Here the scalar bubble graph is given by (\ref{bubbs}), and
\begin{eqnarray}
\Pi_s(z,q)=6i \int \frac{{\rm d}^4 k}{(2\pi)^4} \delta\left(z-\frac{k_-}{q_-}\right)
{\rm tr}_D \left[\gamma_5 \, S(k) \gamma_5 \, S(-(q-k)) \right]\,. \label{bubz}
\end{eqnarray}
Here we use again the relations $S(k)=S_0(k_{\rm Q}),
\, S(-(q-k))=S_0(-(q-k)_{\rm Q})=S_0(k_{\rm Q}-q_{\rm D})$.
Then we perform a shift
$k \rightarrow k+V$, and rewrite the delta function as
\begin{eqnarray}
\delta\left(z-\frac{k_- + V_-}{q_-}\right) = \frac{q_-}{q_{{\rm D}-}} \delta \left(z' - \frac{k_-}{q_{{\rm D}-}}\right)\,,
\end{eqnarray}
where ${\displaystyle z' \equiv \frac{q_-}{q_{{\rm D}-}}z -
\frac{V_-}{q_{{\rm D}-}}}$. Noting also that
$\Pi_s(q)=\Pi_{s0}(q_{\rm D})$,
we obtain
\begin{eqnarray}
f_{q/D}(z) =  \frac{q_-}{q_{{\rm D}-}}\, f_{q/D0}\left(z'= \frac{q_-}{q_{{\rm D}-}}z - \frac{V_-}{q_{{\rm D}-}}\right)\,.
\label{fqd}
\end{eqnarray}
The distributions (\ref{fdn}) and (\ref{fqd}) must then be inserted into the convolution (\ref{convd}). Here we
must note that, because of the presence of the vector field, the distribution (\ref{fqd}) actually depends not
only on $z$, but only on $q_-$, or on $y$ (since $y=q_-/p_-$).
Convolution integrals of this type are treated generally
in Appendix C, and here we have a special case of (\ref{con}) for
$h=D$ and $H=N$.
Therefore we can immediately read off the result from (\ref{res}):
\begin{eqnarray}
f_{q/N}^{(D)}(x) = \frac{p_-}{p_{{\rm N}-}}\, f_{q/N0}^{(D)}(x'=\frac{p_-}{p_{{\rm N}-}}x - \frac{V_-}{p_{{\rm N}-}})
\,,  \label{dd}
\end{eqnarray}
where the distribution $f_{q/N0}^{(D)}(x')$ differs from the free nucleon
expression only by the replacement of the free masses by the effective ones.
The sum of (\ref{qq}) and (\ref{dd}) then reproduces the more general relation (\ref{relation}) of the
main text.

\section{Spectral representations of LC momentum distributions}
\setcounter{equation}{0}
In this Appendix we show that the support of $f_{q/H}(z)$ given by Eq.(\ref{supq1}) is
consistent with the spectral representation (\ref{spq}).
Following the discussion of Ref.\cite{JAFNP}, the contributions to the spectral sum on
the r.h.s. of (\ref{spq}) can be represented by the four diagrams of Fig.8 \footnote{
In general, the arrows in these diagrams denote the flow of quark (or baryon) number:
For example, the left vertex in diagram 1 means to remove a quark with momentum
$p_-\,x$, or to add an antiquark with momentum $-p_-\,x$, and vice versa for the right vertex.
However, since the calculations of the main part are restricted to the valence quark
picture, we need not consider the aniquarks here.}.

\begin{figure}[h]
\begin{center}
\epsfig{file=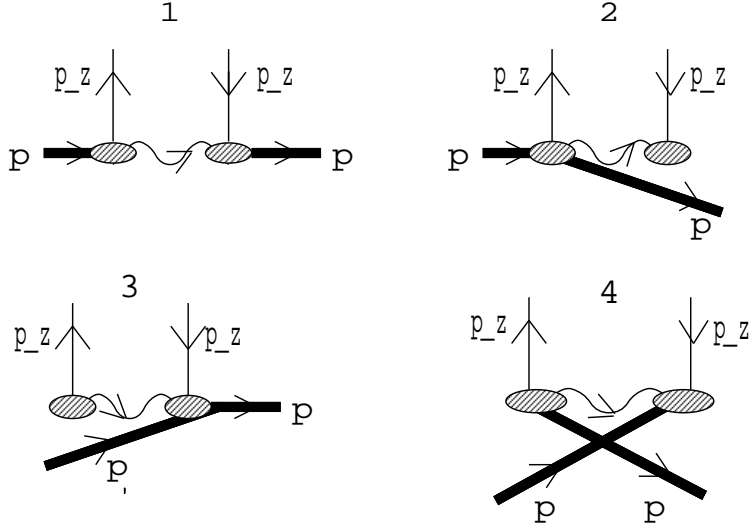,angle=0,width=10cm}
\caption{Diagrammatic representation of the r.h.s. of the spectral representation (\ref{spq}). The
bold line denotes the
external hadron H, the thin line a quark, and the wavy line the
intermediate state.}
\end{center}
\end{figure}

In all four diagrams, we have from (\ref{spq})
\begin{equation}
p_-^{(n)} = p_-(1-z)\,.  \label{mc}
\end{equation}
In the following we discuss the support from these diagrams separately:
\begin{enumerate}
\item Diagram 1: The intermediate state $|n>$ has quark number $(H-1)$, where $H$ is the quark number of the
initial hadron H. Because the minus component of the kinetic momentum must be positive
definite, we obtain
\begin{eqnarray}
p_-^{(n)} - (H-1)V_- = p_-(1-z) - (H-1)V_- >0\,,  \nonumber
\end{eqnarray}
which gives the support
\begin{eqnarray}
z<1-\frac{(H-1)V_-}{p_-}\,\,\,\,\,\,\,\,\,\,({\rm diagram}\,\,1)\,. \label{r1}
\end{eqnarray}
\item Diagrams 2 and 3: The intermediate state is $|n>=|n'>\otimes|H,p>$, where $|n'>$ is indicated by
the wavy line in the diagrams, and has quark number $-1$ and minus momentum component given by
\begin{equation}
p_-^{(n')}=p_-^{(n)}-p_- = -p_-\,z\,, \label{e2}
\end{equation}
where we used (\ref{mc}) in the second step. Since the minus component of the kinetic momentum of the
state $|n'>$ must be positive definite, we obtain
\begin{eqnarray}
p_-^{(n')} +V_-   = -p_-\,z + V_- >0\,,  \nonumber
\end{eqnarray}
which gives the support
\begin{eqnarray}
z < \frac{V_-}{p_-}\,\,\,\,\,\,\,\,\,\,({\rm diagrams}\,\,2\,\,{\rm and}\,\,3)\,. \label{r2}
\end{eqnarray}
\item Diagram 4: The intermediate state is $|n>=|n'>\otimes |H,p>\otimes |H,p>$, where $|n'>$ is indicated by
the wavy line in the diagram, and has quark number $-(H+1)$ and minus momentum component given by
\begin{equation}
p_-^{(n')} = p_-^{(n)}-2p_- = -p_-\,(1+z) \label{e4}
\end{equation}
where we used (\ref{mc}) in the second step. Since the minus component of the kinetic momentum of the
state $|n'>$ must be positive definite, we obtain
\begin{eqnarray}
p_-^{(n')} +(H+1)\,V_-   = -p_-\,(1+z) + (H+1)V_- >0\,, \nonumber
\end{eqnarray}
which gives the support
\begin{eqnarray}
z<-1+\frac{(H+1)V_-}{p_-}\,\,\,\,\,\,\,\,\,\,({\rm diagram}\,\,4) \label{r4}
\end{eqnarray}
\end{enumerate}

We therefore see \footnote{Note that the r.h.s. of (\ref{r4}) is smaller than ${\displaystyle \frac{V_-}{p_-}}$,
because the minus component of the kinetic momentum of the external hadron H
is positive: $p_--HV_->0$.} that in the region
\begin{eqnarray}
\frac{V_-}{p_-} < z < 1 - \frac{(H-1)V_-}{p_-}   \label{rq}
\end{eqnarray}
only diagram 1 contributes. Since diagram 1 corresponds to the ``quark distribution'' in the parton model
sense, we obtain the support as given by Eq. (\ref{supq}).

\section{Convolution integrals}
\setcounter{equation}{0}
Here we consider convolution integrals of the form
\begin{eqnarray}
f_{q/H}(x)=\int {\rm d} y \int {\rm d} z\,\delta(x-yz)\,f_{q/h}(z)\, f_{h/H}(y) \,, \label{con}
\end{eqnarray}
where the distributions refer to hadrons H and h with quark numbers $H$ and $h$, respectively, and the dependence
on the mean vector field is expressed as
\begin{eqnarray}
f_{h/H}(y)&=&\frac{P_-}{P_{{\rm H}-}} \, f_{h/H0}\left(y'=\frac{P_-}{P_{{\rm H}-}}y-\frac{h V_-}{P_{{\rm H}-}}\right)
\label{f1} \\
f_{q/h}(z)&=&\frac{p_-}{p_{{\rm h}-}} \, f_{q/h0}\left(z'=\frac{p_-}{p_{{\rm h}-}}z-\frac{V_-}{p_{{\rm h}-}}\right)\,. \label{f2}
\end{eqnarray}
The minus momentum components of the hadronic states H and h are $P_-$ and $p_-$, respectively, and their kinetic
momenta are defined as in (\ref{kin}). For the case of the convolution
(\ref{conv}), which is needed in the main text, we have to identify H=A, h=N, and $P_{-} \equiv \epsilon_F/\sqrt{2}$,
$P_{{\rm H}-} \equiv E_F/\sqrt{2}$.

To evaluate (\ref{con}), we make the transformation of variables
\begin{eqnarray}
(y, z) &\rightarrow& (p_-, k_-), \,\,\,\,\,\,\,\,\,\,{\rm where} \nonumber \\
y &\equiv& \frac{p_-}{P_-} ; \,\,\,\,\,\,\,\,\,\,z\equiv \frac{k_-}{p_-}\,, \label{t}
\end{eqnarray}
and obtain from the Jacobian of this transformation
\begin{eqnarray}
{\rm d}y\, {\rm d}z = \frac{1}{P_-\,p_-} {\rm d}p_-\, {\rm d}k_-\,.  \label{t1}
\end{eqnarray}
In terms of these variables we have $x=k_-/P_-$, and the quantities $y'$ and $z'$ in (\ref{f1}) and (\ref{f2})
are expressed as
\begin{eqnarray}
y' = \frac{p_{{\rm h}-}}{P_{{\rm H}-}} \,\,\,\,\,\,\,\,\,\,\,\,z' = \frac{k_{{\rm Q}-}}{p_{{\rm h}-}}\,. \label{pr}
\end{eqnarray}
Then Eq.(\ref{con}) becomes
\begin{eqnarray}
f_{q/H}(x) &=& \int {\rm d}k_- \int {\rm d} p_- \,\delta\left(x-\frac{k_-}{P_-}\right)
\frac{1}{P_{{\rm H}-}\,p_{{\rm h}-}}
f_{h/H0}\left(\frac{p_{{\rm h}-}}{P_{{\rm H}-}}\right)\,f_{q/h0}\left(\frac{k_{{\rm Q}-}}{p_{{\rm h}-}}\right) \nonumber \\
\label{c1}
&=&
\int {\rm d}k_{{\rm Q}-} \int {\rm d} p_{{\rm h}-} \, \delta\left(x-\frac{k_{{\rm Q}-}+V_-}{P_{{\rm H}-}+HV_-}\right)
\frac{1}{P_{{\rm H}-}\,p_{{\rm h}-}}
\nonumber \\
&\times& f_{h/H0}\left(\frac{p_{{\rm h}-}}{P_{{\rm H}-}}\right)\,f_{q/h0}\left(\frac{k_{{\rm Q}-}}{p_{{\rm h}-}}\right)\,,
\label{c2}
\end{eqnarray}
where we performed a shift $(k_-, p_-) \rightarrow (k_{{\rm Q}-}, p_{{\rm h}-})$ in the second step.
We then introduce the variables
\begin{eqnarray}
y' &=& \frac{p_{{\rm h}-}}{P_{{\rm H}-}}, \,\,\,\,\,\,\,\,\,\,z'=\frac{k_{{\rm Q}-}}{p_{{\rm h}-}} \nonumber \\
&\Rightarrow& p_{{\rm h}-} = P_{{\rm H}-}\,y', \,\,\,\,\,\,\,\,\,\,k_{{\rm Q}-} = P_{{\rm H}-}\,y'z' \label{tt} \\
&\Rightarrow& \, \delta\left(x-\frac{k_{{\rm Q}-}+V_-}{P_{{\rm H}-}+HV_-}\right)
= \frac{P_-}{P_{{\rm H}-}} \, \delta\left(\frac{P_-}{P_{{\rm H}-}}x - z'y' - \frac{V_-}{P_{{\rm H}-}}\right) \nonumber \\
&\equiv& \frac{P_-}{P_{{\rm H}-}} \, \delta\left(x' - z'y' \right)\,,  \label{d}
\end{eqnarray}
where in the last step we defined
\begin{eqnarray}
x' \equiv \frac{P_-}{P_{{\rm H}-}}x - \frac{V_-}{P_{{\rm H}-}}\,. \label{xp}
\end{eqnarray}
Then we obtain finally
\begin{eqnarray}
f_{q/H}(x)&=&\frac{P_-}{P_{{\rm H}-}} \int {\rm d} y' \int {\rm d} z'\,\delta(x'-y'z')\,f_{h/H0}(y') f_{q/h0}(z') \label{con0}
\nonumber \\
&=& \frac{P_-}{P_{{\rm H}-}} f_{q/H0}\left(x' = \frac{P_-}{P_{{\rm H}-}}x - \frac{V_-}{P_{{\rm H}-}}\right)\,. \label{res}
\end{eqnarray}

\section{Moments and distribution functions in the proper time regularization scheme}
\setcounter{equation}{0}
In this Appendix we derive the quark distribution functions in the absence of the vector field given in
Subsect.3.2.

The contribution of the quark diagram of Fig.1 to the moments is given by (\ref{anq}), where only the
pole term of $\tau_0$ (see Eq.(\ref{pole})) contributes. We combine the two denominators by a Feynman
parameter $\alpha$, and shift the loop momentum according to $k\rightarrow k+p\alpha$ to obtain
\begin{eqnarray}
A_n^{(Q)} = i\,g_D \, {\overline \Gamma}_N(p) \frac{\partial}{\partial p_+}
\int \frac{{\rm d}^4k}{(2\pi)^4}\,\int_0^1 {\rm d} \alpha
\left(\frac{k_- + p_- \alpha}{p_-}\right)^{n-1} 
\frac{\fslash{k}+\fslash{p}\alpha + M}{\left(k^2 - A(\alpha,p^2)\right)^2}\, \Gamma_N(p)\,,   
\nonumber
\end{eqnarray}
where $A(\alpha,p^2)$ is defined in Eq.(\ref{a}).
In general, invariant integration over a function of the form ${\displaystyle f(k^2) k^{\alpha} k^{\beta} \dots}$
gives a sum of products of contractions $g^{\alpha \beta}\dots$. If we note that $g^{--}=0$, it is easy to see
that, in the expansion of $\left(k_-+p_- \alpha\right)^{n-1}$, all terms $\propto (k_-)^m$
give no contributions if $m\geq 2$. We can therefore replace
$\left(k_-+p_- \alpha\right)^{n-1} \rightarrow \left(p_- \alpha\right)^{n-1} + (n-1) k_- \left(p_- \alpha\right)^{n-2}$.
If we further use ${\displaystyle \fslash{k}=k_+ \gamma_- + k_- \gamma_+ - {\bold k}_{\perp}\cdot {\bold \gamma}_{\perp}}$,
as well as the form (\ref{gamma}) of the nucleon spinor, we obtain
\begin{eqnarray}
A_n^{(Q)} &=& - i g_D\,Z_N\,
\frac{M_N}{p_-} {\overline u}_N(p) \left[\frac{\partial}{\partial p_+}
\int \frac{{\rm d}^4k}{(2\pi)^4}\,
\int_0^1 {\rm d} \alpha\, \alpha^{n-1} \right. \nonumber \\
&\times& \left. \frac{\alpha \fslash{p} + M + \frac{n-1}{2 \alpha p_-} \left(k_0^2 - k_3^2 \right) \, \gamma_-}
{\left(k^2 - A(\alpha,p^2)\right)^2} \right] u_N(p)\,. \nonumber
\end{eqnarray}
By using the Dirac equation,
we can replace $\gamma_- \rightarrow p_-/M_N$ in the last term of the above
expression. Because of $p^2={\fslash p}^2$,
the integrand can then be considered as a function
$F({\fslash p})$, and by noting that
\begin{eqnarray}
\frac{M_N}{p_-} \overline{u}_N(p) \left(\frac{\partial}{\partial p_+}\,F(\fslash{p}) \right) u_N(p)
=
\left(\frac{\partial F}{\partial {\fslash p}}\right)_{\fslash{p}=M_N}  \nonumber
\end{eqnarray}
we arrive at the form (\ref{anq1}) given in the main text. If we perform a partial integration 
in $\alpha$ and note that the surface term is independent of $\fslash{p}$, we obtain the
expression (\ref{one}) for the distribution function.

We perform a Wick rotation in (\ref{one}), and note that in any $O(4)$ invariant Euclidean
regularization scheme we can replace $k_0^2+k_3^2 \rightarrow k^2/2$, where now our notation
refers to Euclidean metric ($k^2=k_0^2+{\bold k}^2$). By introducing the proper time regularization
scheme (\ref{pt}), we obtain
\begin{eqnarray}
f_{q/N0}^{(Q)}(x) = g_D\,Z_N\, \frac{\partial}{\partial {\fslash p}}
\int \frac{{\rm d}^4k}{(2\pi)^4}\, \int_{1/\Lambda_{\rm UV}^2}^{1/\Lambda_{\rm IR}^2}
\, \tau\,{\rm d}\tau\, \left( \fslash{p} x + M + \frac{k^2}{4 M_N} \frac{\rm d}{{\rm d}x} \right)
e^{-\tau\left(k^2+A(x,p^2)\right)} \nonumber
\end{eqnarray}
It is then easy to perform the loop integral and the required derivatives to arrive at the
result (\ref{fxq}) of the main text. 

Next we consider the quark LC momentum distribution in the diquark. The moments are given by Eq.(\ref{anqd}).
We calculate the Dirac trace, introduce a Feynman parameter $\alpha$, and shift the loop momentum
according to $k \rightarrow k + q \alpha$ to obtain
\begin{eqnarray}
A_n^{(q/D)} &=& - 24i\,g_D\,\frac{\partial}{\partial q^2} \int \frac{{\rm d}^4k}{(2\pi)^4} \int_0^1 {\rm d}\alpha\,
\left(\frac{k_- + q_- \alpha}{q_-} \right)^{n-1} \nonumber \\ 
&\times& \frac{M^2+q^2\alpha(1-\alpha) - k^2 +  k \cdot q(1-2\alpha)}{\left(k^2 +q^2 \alpha (1-\alpha) - M^2 \right)^2}\,, 
\nonumber
\end{eqnarray}
where one has to set $q^2=M_D^2$ at the end. Following the same argument as given above for the quark
diagram, there are no contributions from $(k_-)^m\,\,(m \geq 2)$ in the expansion of $(k_- + q_- \alpha)^{n-1}$
because of $g^{--}=0$.
By using $k \cdot q = k_+ q_+ + k_- q_- - {\bold k}_{\perp} \cdot {\bold q}_{\perp}$, we obtain
the expression (\ref{anqd1}) of the main text. Performing a partial integration in $\alpha$ and noting
that the surface term is independent of $q^2$ leads to the distribution function (\ref{onep}).
We then perform a Wick rotation in (\ref{onep}), and note again that in any $O(4)$ invariant Euclidean
regularization scheme we can replace $k_0^2+k_3^2 \rightarrow k^2/2$, where now 
$k^2=k_0^2+{\bold k}^2$. By introducing the proper time regularization
scheme (\ref{pt}), we obtain
\begin{eqnarray}
f_{q/D0}(x) &=& 24 \, g_D\, \frac{\partial}{\partial q^2} \int \frac{{\rm d}^4k}{(2\pi)^4} 
\int_{1/\Lambda_{\rm UV}^2}^{1/\Lambda_{\rm IR}^2} \, {\rm d}\tau \nonumber \\ 
&\times& \left( 1 + 2 q^2 x(1-x) \tau + \frac{k^2 \tau}{4} \frac{\rm d}{{\rm d}x} (1-2x) \right) 
e^{-\tau \left(k^2+ M^2 - q^2 x(1-x)\right)}\,, \nonumber
\end{eqnarray}
where the derivative ${\rm d}/{\rm d}x$ acts on the whole function on its r.h.s. 
It is then easy to perform the loop integral and the required derivatives to arrive at the
result (\ref{fxd}) of the main text.

\section{Results for the case ${\bold \rho=}${\bf 0.16 fm}$^{\bf -3}$}
\setcounter{equation}{0}
The numerical calculations discussed in Sect.4 were performed at the
saturation density in our model
($\rho=0.22$ fm$^{-3}$),
which is too high compared to the empirical value. In order to indicate
the dependence on the density,
we will show the results for $\rho=0.16$ fm$^{-3}$ in this Appendix,
using the effective masses and energies listed in Table 1.
One should keep in mind, however,
that in the derivations of the main text, for example
of Eqs.(\ref{relation0}) and
(\ref{fqa1}), the relation ${\overline M}_N=\epsilon_F$ was assumed. As
this is satisfied only at the saturation density
(see Table 1),
the results shown in this Appendix should serve only to
indicate qualitatively the density dependence
of the results.

Fig.9 shows how the results of Fig. 5 change if we decrease the density to $\rho=0.16$ fm$^{-3}$.

\begin{figure}[h]
\begin{center}
\epsfig{file=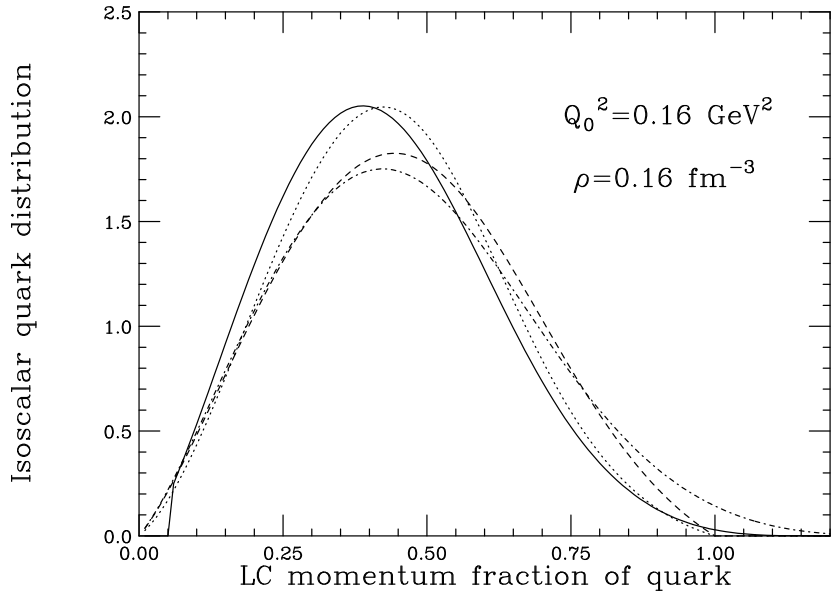,angle=90,width=11cm}
\caption{Same as Fig.5 for the density $\rho=0.16$ fm$^{-3}$.}
\end{center}
\end{figure}

We see that the various medium modifications of the quark distribution function show a very similar
behavior, but their magnitudes are smaller because of the smaller density.
The structure function per nucleon in nuclear matter (see Eq.(\ref{f2a})) is compared to the free nucleon
structure function in Fig. 10. In order
to plot the function (\ref{f2a}) against $x$, we used the relation (\ref{bja}) with
${\overline M}_N$ equal to the mass per nucleon at the density $\rho=0.16$ fm$^{-3}$ given in Table 1.

\begin{figure}[h]
\begin{center}
\epsfig{file=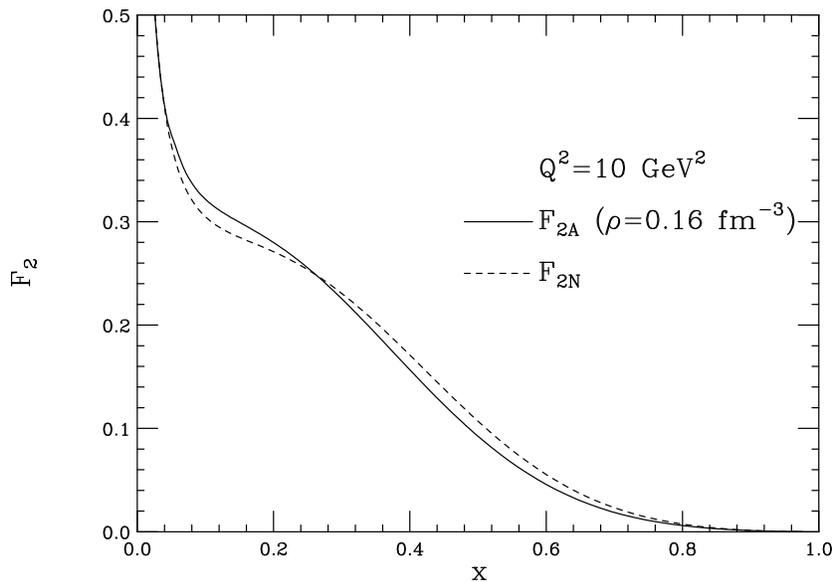,angle=90,width=11cm}
\caption{Same as Fig.6 for the density $\rho=0.16$ fm$^{-3}$.}
\end{center}
\end{figure}

The qualitative behavior is the same as in Fig. 6, but the difference between the dashed and solid lines
is smaller. Finally, the ratio (\ref{emcn}) is shown in
Fig.11. Compared to Fig.7, the magnitude of the EMC effect is reduced. 

\begin{figure}[h]
\begin{center}
\epsfig{file=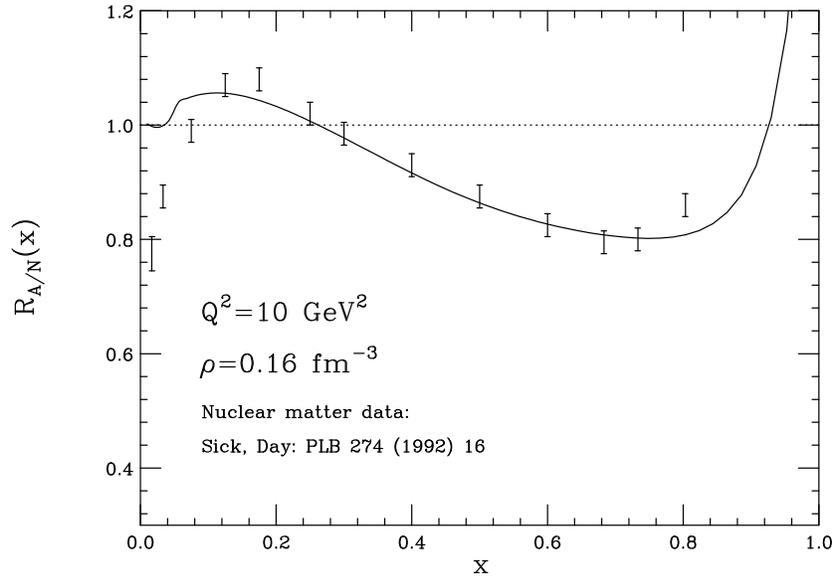,angle=90,width=11cm}
\caption{Same as Fig.7 for the density $\rho=0.16$ fm$^{-3}$.}
\end{center}
\end{figure}

\end{document}